\documentclass[%
aip,
longbibliography,
jcp,%
amsmath,amssymb,
reprint%
]{revtex4-1}
\usepackage[latin1]{inputenc}

\usepackage{amsmath}
\usepackage{amsfonts}
\usepackage{amssymb,bm}
\usepackage{graphicx}
\usepackage{physics}
\usepackage{upgreek}
\usepackage[outercaption]{sidecap}   
\usepackage{color}
\usepackage[normalem]{ulem}
\usepackage{soul}
\mathchardef\mhyphen="2D
\usepackage{txfonts}
\usepackage{subcaption}
\usepackage{hyperref}
\hypersetup{colorlinks=true, citecolor=blue, urlcolor=blue, linkcolor=blue}

\DeclareMathAlphabet{\pazocal}{OMS}{zplm}{m}{n}

\usepackage{xr}
\begin{document}

\title{Fast quasi-centroid molecular dynamics for water and ice}
\author{Joseph E. Lawrence}
 \email{joseph.lawrence@phys.chem.ethz.ch}
\affiliation{Laboratory of Physical Chemistry, ETH Z\"urich, 8093 Z\"urich, Switzerland}

\author{Annina Z. Lieberherr}%
\affiliation{ Physical and Theoretical Chemistry Laboratory, Department of Chemistry, University of Oxford, South Parks Road, Oxford OX1 3QZ, United Kingdom}%

\author{Theo Fletcher}
\affiliation{ Physical and Theoretical Chemistry Laboratory, Department of Chemistry, University of Oxford, South Parks Road, Oxford OX1 3QZ, United Kingdom}%

\author{David E. Manolopoulos}
\affiliation{ Physical and Theoretical Chemistry Laboratory, Department of Chemistry, University of Oxford, South Parks Road, Oxford OX1 3QZ, United Kingdom}%

\date{\today}

\begin{abstract}
{\bf \small \textsf{ABSTRACT:}} We describe how the fast quasi-centroid molecular dynamics (f-QCMD) method can be applied to condensed phase systems by approximating the quasi-centroid potential of mean force as a sum of inter- and intra-molecular corrections to the classical interaction potential.  The corrections are found by using a regularised iterative Boltzmann inversion procedure to recover the inter- and intra-molecular quasi-centroid distribution functions obtained from a path integral molecular dynamics simulation. The resulting methodology is found to give good agreement with a previously published QCMD dipole absorption spectrum for liquid water, and satisfactory agreement for ice. It also gives good agreement with spectra from a recent implementation of CMD that uses a pre-computed elevated temperature potential of mean force. Modern centroid molecular dynamics methods therefore appear to be reaching a consensus regarding the impact of nuclear quantum effects on the vibrational spectra of water and ice.
\end{abstract}

\maketitle

\section{Introduction}

The importance of experimental vibrational spectra as a probe of inter- and intra-molecular interactions has generated considerable interest in the development of methods for their simulation. For small gas phase molecules it is possible to compute vibrational spectra exactly using wavefunction methods. However, for condensed phase systems like liquid water and ice,  it is not feasible to solve the Schr\"odinger equation exactly due to the exponential scaling of quantum mechanics with dimensionality. It is nevertheless possible to compute the static equilibrium properties of condensed phase systems using imaginary-time path-integral techniques, which are based on the isomorphism between distinguishable quantum particles and classical ring polymers.\cite{Chan1981} Indeed the path integral molecular dynamics (PIMD) method has been used for almost forty years\cite{Pari1984} to shed light on quantum mechanical (zero point energy and tunnelling) effects in condensed phases. There has therefore been  particular interest in adapting methods based on imaginary time path integrals to provide approximations to the dynamical properties of condensed phase systems, such as their vibrational spectra.

In a pioneering series of papers published almost thirty years ago, Cao and Voth focussed on the centroid of the ring polymer and showed how this could be used to develop such a dynamical approximation.\cite{Cao1994a,Cao1994b,Cao1994c,Cao1994d} Their centroid molecular dynamics (CMD) method is simply classical molecular dynamics on an effective potential: the potential of mean force obtained by averaging the thermal fluctuations of the ring polymer around its Cartesian centroid. 
Various approximations to this effective potential had been suggested previously,\cite{Feyn1965,Feyn1986} but to keep their method consistent with static equilibrium properties, Cao and Voth chose not to make any approximation to the quantum mechanical partition function.\cite{Cao1994a} Instead, they developed an adiabatic algorithm that enabled them to calculate the exact mean force on the centroid `on the fly' during the course of a PIMD simulation.\cite{Cao1994d} This adiabatic algorithm has since been used to apply CMD to a wide range of systems and to study the impact of nuclear quantum effects on a variety of dynamical observables, including vibrational spectra.\cite{Paes2009,Paes2010} 

The resulting vibrational spectra are generally quite reliable at high temperatures, and the method is expected to give a good approximation to the quantum transition frequency $\omega_{10}=(E_1-E_0)/\hbar$ of a one-dimensional anharmonic oscillator in the low temperature limit.\cite{Rami1998} However, CMD has since been found to suffer from a `curvature problem' in multidimensional systems that possess both stretching and angular (bending, torsional, or librational) modes.\cite{Witt2009} The ring polymer spreads out around the angular coordinates as the temperature is lowered, causing a softening of the centroid potential of mean force in the radial directions that results in a spurious red-shifting and broadening of the stretching bands in the CMD spectrum.\cite{Witt2009} While the shifting and broadening of the O--H stretching band are not especially pronounced in liquid water at room temperature,\cite{Ivan2010,Paes2010} the problem becomes more severe as the temperature is lowered, and it is certainly quite noticeable in ice at 150 K.\cite{Bens2020}

To solve the curvature problem, Althorpe and co-workers have introduced a `quasi-centroid' molecular dynamics (QCMD) method in which the dynamical variable is the quasi-centroid of the ring polymer rather than its centroid.\cite{Tren2019} This quasi-centroid is defined in terms of certain radial and angular coordinates that are specific to the problem at hand. For example, the quasi-centroid O--H bond lengths of a water molecule are
\begin{equation}\label{eq:rbar}
 {r}_i = {1\over P}\sum_{j=1}^P r_i^{(j)}
\end{equation}
and the quasi-centroid H--O--H bond angle is
\begin{equation}\label{eq:thetabar}
 {\theta} = {1\over P}\sum_{j=1}^P \theta^{\,(j)},
\end{equation}
where $r_1^{(j)}$ and $r_2^{(j)}$ are the two O--H bond lengths and $\theta^{\,(j)}$ is the H--O--H bond angle of the molecule in the $j$-th bead of the ring polymer necklace. Replacing the centroid with the quasi-centroid solves the curvature problem because the radial quasi-centroids $ {r}_i$ remain at realistic bond lengths when the temperature is lowered and the bead distributions in the rotational and bending coordinates become less compact. 

Althorpe and co-workers have shown that QCMD vibrational spectra have physically reasonable stretching bands at both high and low temperatures for systems including gas phase water\cite{Tren2019} and ammonia,\cite{Hagg2021} and condensed phase water and ice.\cite{Bens2020}
These QCMD calculations were performed using a quasi-centroid adaptation\cite{Tren2019} of Cao and Voth's adiabatic CMD algorithm.\cite{Cao1994d} 
Unfortunately, this is even more computationally expensive than the original adiabatic CMD algorithm. One has to use a small time step to correctly integrate the equations of motion of the adiabatically separated ring polymer internal modes and the fact that these modes are coupled to the quasi-centroid leads to additional complications that have to be dealt with rather carefully.\cite{Tren2022} 

To avoid these difficulties, we have recently developed a `fast' implementation of QCMD in which one pre-computes an approximation to the quasi-centroid potential of mean force that is consistent with the quasi-centroid distribution functions obtained from a PIMD simulation.\cite{Flet2021} The idea here is similar to that of an earlier `fast CMD' method of Hone, Izvekov and Voth,\cite{Hone2005} and in both cases the potential of mean force is calculated by adding a small correction to the classical interaction potential. However, while Hone {\em et al.} used force matching to fit their correction to PIMD forces, we have found it more convenient in the quasi-centroid case to use iterative Boltzmann inversion\cite{Sope1996,Reit2003} (IBI) to fit the correction to PIMD distribution functions. The resulting methodology reproduces Althorpe and co-worker's QCMD results for gas phase water and ammonia, and it enabled us to perform the first QCMD calculations for gas phase methane.\cite{Flet2021} In Sec.~II we will describe how fast QCMD can be generalised from the gas phase to the condensed phase and used to simulate the vibrational spectra of liquid water and ice.

More recently still, Kapil and co-workers have developed an alternative method in which the centroid is retained in preference to the quasi-centroid, but the centroid potential of mean force is calculated at an elevated temperature.\cite{Musi2022} This is chosen to be sufficiently high that the curvature problem has not yet become an issue, and yet sufficiently low that the high frequency intra-molecular stretching vibrations are already in their quantum ground states. A suitable temperature for liquid water can be found by monitoring the convergence of the PIMD distribution function in the O--H stretching coordinate as the temperature is lowered. Since this stops changing at around 600~K, where the curvature problem is not yet an issue, a centroid potential of mean force constructed at this temperature can be used in simulations of the vibrational spectrum at lower temperatures. Kapil and co-workers employed machine learning techniques to construct the 600~K centroid potential of mean force and demonstrated that the resulting `elevated temperature path integral ground state' (Te PIGS) method avoids the curvature problem in liquid water at 300~K and in ice at lower temperatures.\cite{Musi2022} We shall compare our f-QCMD results with theirs for both of these systems in Sec.~III.

\section{Fast QCMD}

A fast QCMD simulation of a system such as liquid water or ice has three stages: (i) a short PIMD simulation is used to construct quasi-centroid distribution functions at the target temperature and density; (ii) an IBI (coarse-graining) procedure is used to fit these distribution functions to those of an effective classical potential of mean force; (iii) a classical simulation of the vibrational spectrum is performed on the resulting quasi-centroid potential of mean force.
We shall now describe each of these stages in turn.

\subsection{Quasi-centroid distribution functions}

We have already defined what we mean by the quasi-centroid bond lengths and the quasi-centroid bond angle of a given water molecule at a particular configuration in a $P$-bead PIMD simulation [Eqs.~\eqref{eq:rbar} and \eqref{eq:thetabar}]. These three coordinates, $ {r}_1$, $ {r}_2$ and $ {\theta}$, suffice to determine the size and shape of the quasi-centroid water molecule, but not its position or orientation in space. In order to calculate inter-molecular quasi-centroid distribution functions, we need to know the actual positions $ {\bm r}_{\rm X}$ of the quasi-centroids of all three atoms (X = O, H$_1$, and H$_2$) in the molecule.
To determine them, we follow Althorpe and co-workers in setting the centre-of-mass of the quasi-centroid molecule to be that of the Cartesian centroid molecule, and choosing the orientation of the quasi-centroid molecule to be as close as possible to that of the Cartesian centroid molecule.\cite{Tren2019,Tren2022} Given an arbitrarily aligned and positioned quasi-centroid molecule constructed from $ {r}_1$, $ {r}_2$, and $ {\theta}$, we first find the rotation matrix ${\bf U}$ that minimises the sum of mass-weighted squared deviations\cite{Kras2014}
\begin{equation}
w({\bf U}) = \sum_{\rm X} m_{\rm X}\left(\left[{\bm r}^{({\rm c})}_{\rm X}-{\bm r}^{({\rm c})}_{\rm CM}\right]-{\bf U}\left[{\bm r}^{({\rm qc})}_{\rm X}-{\bm r}^{({\rm qc})}_{\rm CM}\right]\right)^2
\end{equation}
where ${\bm r}^{({\rm c})}_{\rm X}$ and ${\bm r}^{({\rm qc})}_{\rm X}$ are the centroid and quasi-centroid of atom X and ${\bm r}^{({\rm c})}_{\rm CM}$ and ${\bm r}^{({\rm qc})}_{\rm CM}$ are the corresponding molecular centres of mass. The translated and rotated quasi-centroid coordinates are then 
\begin{equation}
 {\bm r}_{\rm X}={\bm r}^{({\rm c})}_{\rm CM}+{\bf U}\left[{\bm r}^{({\rm qc})}_{\rm X}-{\bm r}^{({\rm qc})}_{\rm CM}\right].
\end{equation}
This procedure is easier than it may seem because there exists a simple and efficient algorithm for finding the optimum ${\bf U}$ in Eq.~(3) and performing the rotation in Eq.~(4), as described in Ref.~\onlinecite{Kras2014}. Applying this algorithm to each water molecule  in turn rapidly generates a set of quasi-centroid coordinates $\bigl\{ {\bm r}_{{\rm X}I}\bigr\}$ for each atom X within each molecule $I$ in any given configuration of the PIMD simulation.

Armed with these quasi-centroid coordinates, it is straightforward to calculate inter-molecular quasi-centroid radial distribution functions from the configurations visited in the PIMD simulation. These radial distribution functions are defined as
\begin{subequations}\label{eq:grs}
\begin{equation}
g_{\rm OO}( {r}) = {1\over 4\pi r^2N\rho}
\sum_{I=1}^N\sum_{J\not=I}^N \left<\delta\Bigl( {r}-| {\bm r}_{{\rm O}I}- {\bm r}_{{\rm O}J}|\Bigr)\right>,
\end{equation}
\begin{equation}
g_{\rm OH}( {r}) = {1\over 8\pi r^2N\rho}
\sum_{I=1}^N\sum_{J\not=I}^N\sum_{j=1}^2 \left<\delta\Bigl( {r}-| {\bm r}_{{\rm O}I}- {\bm r}_{{\rm H}_jJ}|\Bigr)\right>,
\end{equation}
\begin{equation}
g_{\rm HH}( {r}) = {1\over 16\pi r^2N\rho}
\sum_{I=1}^N\sum_{J\not=I}^N\sum_{i,j=1}^2 \left<\delta\Bigl( {r}-| {\bm r}_{{\rm H}_iI}- {\bm r}_{{\rm H}_jJ}|\Bigr)\right>,
\end{equation}
\end{subequations}
where the angular brackets denote averages over the path integral configurations that are used to generate the quasi-centroid coordinates ${\bm r}_{{\rm X}I}$, and $\rho=N/V$ is the bulk number density of water molecules. All three radial distribution functions are straightforward to accumulate as histograms during a short PIMD simulation.

To complete the information that is used to construct our approximation to the quasi-centroid potential of mean force, we use the same form for the intra-molecular distribution functions as we used in our fast QCMD study of the water monomer.\cite{Flet2021} These are
\begin{equation}\label{eq:rhor}
\rho_r( {r}) = {1\over 2N}\sum_{I=1}^N\sum_{i=1}^2 \left<\delta\Bigl( {r}- {r}_{Ii}\Bigr)\right>,
\end{equation}
and
\begin{equation}\label{eq:rhotheta}
\rho_{\theta}( {\theta}) = {1\over N}\sum_{I=1}^N \left<\delta\Bigl( {\theta}- {\theta}_I\Bigr)\right>,
\end{equation}
where $ {r}_{Ii}$ and $ {\theta}_I$ are the path integral bead averages of the intra-molecular O--H$_i$ distance and the intra-molecular H--O--H bond angle in molecule $I$ [see  Eqs.~\eqref{eq:rbar} and~\eqref{eq:thetabar}]. These distribution functions can be accumulated as histograms during the same PIMD simulation that is used to calculate $g_{\rm OO}( {r})$, $g_{\rm OH}( {r})$, and $g_{\rm HH}( {r})$.

\subsection{Quasi-centroid potential of mean force}

The central observation behind fast QCMD is that, by definition, the quasi-centroid distribution functions in Eqs.~\eqref{eq:grs}, \eqref{eq:rhor} and \eqref{eq:rhotheta} will be the same whether the angular brackets $\left\langle \cdot \right\rangle$ indicate an average over a classical NVT simulation on the quasi-centroid potential of mean-force, or over a PIMD NVT simulation in which the locations of the quasi-centroids $ {\bm r}_{{\rm X}I}$ are calculated from the path-integral beads. This equivalence allows us to use standard coarse-graining techniques such as IBI to construct an approximation to the quasi-centroid potential of mean force that is consistent with the distribution functions obtained from the PIMD simulation.\cite{Flet2021}

To do this, we begin as we did for gas phase water by writing the quasi-centroid potential of mean force as a correction to the classical interaction potential, but now with inter- as well as intra-molecular correction terms:
\begin{equation}\label{eq:QCPMF}
V_{\rm qc}( {\bf r}) = V_{\rm cl}( {\bf r})+\Delta V_{\rm intra}( {\bf r})+\Delta V_{\rm inter}( {\bf r}).
\end{equation}
Here $ {\bf r}$ is a configuration vector of the entire quasi-centroid system, $V_{\rm cl}( {\bf r})$ is the classical interaction potential, and $\Delta V_{\rm intra}( {\bf r})$ and $\Delta V_{\rm inter}( {\bf r})$ are the intra- and inter-molecular correction terms. For computational expedience, both in the IBI and in the subsequent molecular simulation on the potential of mean force $V_{\rm qc}( {\bf r})$, we approximate $\Delta V_{\rm intra}( {\bf r})$ as we did in our study of gas phase water\cite{Flet2021}
\begin{equation}\label{eq:Vintra}
\Delta V_{\rm intra}( {\bf r}) \simeq \sum_{I=1}^N\sum_{i=1}^2 \Delta V_r( {r}_{Ii})+\sum_{I=1}^N \Delta V_{\theta}( {\theta}_I),
\end{equation}
and $\Delta V_{\rm inter}( {\bf r})$ as a sum of pairwise contributions,
\begin{equation}\label{eq:Vinter}
\Delta V_{\rm inter}( {\bf r}) \simeq \sum_{I=1}^N\sum_{J\not=I}^N \sum_{{\rm X}\in I}\sum_{{\rm Y}\in J} \Delta V_{\rm XY}\Bigl(\left| {\bm r}_{{\rm X}I}- {\bm r}_{{\rm Y}J}\right|\Bigr).
\end{equation}

\subsection{Iterative Boltzmann inversion}

The corrections in Eqs.~\eqref{eq:Vintra} and~\eqref{eq:Vinter} can be found by IBI.~\cite{Sope1996,Reit2003}
While there are alternative methods that could be applied to this problem,~\cite{Lyub1995,McGr1988} we have found IBI to be the most efficient and robust.
The algorithm can be explained as follows.
One begins by setting the initial corrections $\Delta V_\mathrm{intra}^{(0)}({\bf r})$ and $\Delta V_\mathrm{inter}^{(0)}({\bf r})$ to zero, such that the initial guess for quasi--centroid potential is just the classical potential $V_{\rm cl}({\bf r})$. In the $i^\mathrm{th}$ iteration of the algorithm, the system is propagated on the effective potential $V^{(i)}_{\rm qc}({\bf r})$ obtained by adding the potential corrections $\Delta V_{\mathrm{intra}}^{(i)}({\bf r})$ and $\Delta V_{\mathrm{inter}}^{(i)}({\bf r})$ to $V_{\rm cl}({\bf r})$. The corresponding approximations to the distribution functions $\rho_r^{(i)}(r)$, $\rho_\theta^{(i)}(\theta)$, and $g_{\mathrm{XY}}^{(i)}(r)$ are computed as classical NVT averages with this effective potential. In the basic IBI algorithm, the potential for the next iteration is then found by updating the potential corrections according to 
\begin{subequations}
\label{eq:v-ibi}
\begin{equation}
    \Delta V_{r}^{(i+1)}(r) = \Delta V_{r}^{(i)}(r) - \frac{1}{\beta} \log\left( \frac{\rho^\mathrm{exact}_{r}(r)}{\rho_{r}^{(i)}(r)} \right),
    \label{eq:vintra-ibi-r}
\end{equation}
\begin{equation}
    \Delta V_{\theta}^{(i+1)}(\theta) = \Delta V_{\theta}^{(i)}(\theta) - \frac{1}{\beta} \log\left( \frac{\rho^\mathrm{exact}_{\theta}(\theta)}{\rho_{\theta}^{(i)}(\theta)} \right),
    \label{eq:vintra-ibi-theta}
\end{equation}
\begin{equation}
    \Delta V_{\mathrm{XY}}^{(i+1)}(r) = \Delta V_{\mathrm{XY}}^{(i)}(r) - \frac{1}{\beta} \log\left( \frac{g^\mathrm{exact}_{\mathrm{XY}}(r)}{g_{\mathrm{XY}}^{(i)}(r)} \right),
    \label{eq:vinter-ibi}
\end{equation}
\end{subequations}
where $\beta = 1/k_{\rm B}T$. The exact distribution functions in these updating formulas are pre-computed in a PIMD simulation as described above.

The most expensive part of this algorithm is the calculation of the PIMD distribution functions. The second most expensive part is the calculation of the effective classical distribution functions, $\rho_r^{(i)}(r)$, $\rho_\theta^{(i)}(\theta)$, and $g_{\mathrm{XY}}^{(i)}(r)$, at each IBI iteration. This is facilitated by the recent introduction of variationally-optimised\cite{Cole2021} low-variance force estimators\cite{Rote2020} for classical distrbution functions. With these new estimators, it is possible to calculate the distribution functions with so little effort that IBI is perfectly feasible even when many iterations are required for convergence. The effort of the IBI does not even approach that of the PIMD calculation until the number of iterations exceeds the number of path integral beads, which we have not found to happen in any of our calculations.

In the present calculations we shall follow Althorpe and co-workers\cite{Bens2020} and Kapil and co-workers\cite{Musi2022} in using the qTIP4P/F water model\cite{Habe2009} to study liquid water and ice. This leads to a further simplification in that the intra-molecular part of the qTIP4P/F potential is a sum of O--H bond length and H--O--H bond angle terms of the same form as $\Delta V_{\rm intra}({\bf r})$ in Eq.~\eqref{eq:Vintra}. In practice, we find that it is sufficient to assume that each correction, $\Delta V_r^{(i)}(r)$ and $\Delta V_{\theta}^{(i)}(\theta)$, is a polynomial of the same degree as in the original intra-molecular potential, which implies that we can simply adjust the parameters in the intra-molecular potential rather than update $\Delta V_r^{(i)}(r)$ and $\Delta V_{\theta}^{(i)}(\theta)$ at each IBI iteration.

The corrections to the inter-molecular potentials $\Delta V_{\rm XY}^{(i)}(r)$ are more problematic, however. We have found that in order to obtain a convergent sequence of these corrections, it is essential to regularise the inter-molecular IBI update, as we shall describe next.

\subsection{Regularised IBI}
\label{subsec:theory-regibi}

There are heuristic arguments to suggest that the IBI algorithm will always find a solution,\cite{Sope1996} and progress has recently been made towards proving its convergence.~\cite{Hank2017} In practice, however, we have found (in agreement with others~\cite{Reit2003}) that close to convergence the iterations can start to oscillate, and eventually even become unstable. Furthermore, in regions where the radial distribution functions $g_{\rm XY}(r)$ are small, the IBI update can become dominated by statistical errors.   

We can overcome both problems by adding a regularisation term to the radial distribution functions before performing the IBI update. The new update formula is
\begin{equation}
    \Delta V^{(i+1)}_{\mathrm{XY}}(r) = \Delta V^{(i)}_{\mathrm{XY}}(r) - \frac{1}{\beta} \log\left(\frac{g^{\rm exact}_{\mathrm{XY}}(r)+\varepsilon \,G_{\rm XY}}{g^{(i)}_{\mathrm{XY}}(r)+\varepsilon\,G_{\rm XY}}\right), \label{eq:ibireg}
\end{equation}
where $\varepsilon$ is a non-negative scalar parameter and $G_{\rm XY}$ is the larger of the maximum peak heights in $g_{\mathrm{XY}}^\mathrm{exact}(r)$  and $g_{\mathrm{XY}}^{(i)}(r)$. This definition of $G_{\rm XY}$ allows us to use the same value of $\varepsilon$ for all three radial distribution functions (OO, OH, and HH). The regularisation in Eq.~\eqref{eq:ibireg} was first proposed by Soper,~\cite{Sope1996} but it has  (to the best of our knowledge) not been picked up in later applications of IBI. 
There is no need to regularise the intra-molecular correction because 
only the regions of $\rho_r(r)$ and $\rho_{\theta}(\theta)$ with low  statistical noise need be used in the adjustment of the intra-molecular potential parameters.

\begin{figure}[tb]
\includegraphics[width=.9\columnwidth]{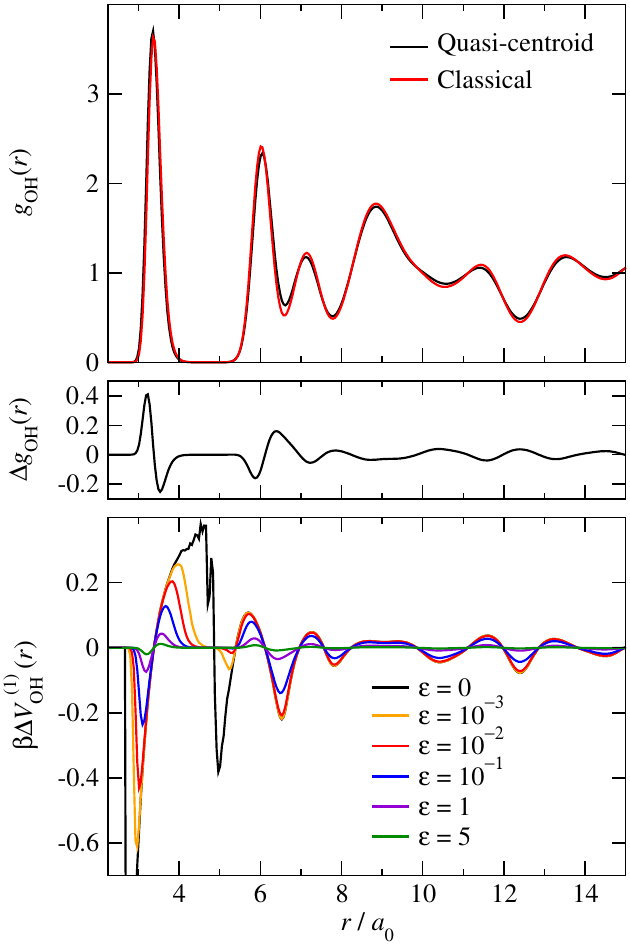}
\caption{Influence of the regularisation parameter $\varepsilon$ on the shape of the first IBI update to the inter-molecular O--H potential in ice at 150\,K. The top panel shows the classical and quasi-centroid radial distribution functions, the middle panel shows the difference between them, and the bottom panel shows the correction potential for different values of $\varepsilon$.}
\label{fig:ibi-varyeps}
\end{figure}

Fig.~\ref{fig:ibi-varyeps} shows the effect of $\varepsilon$ on the first update to the O--H intermolecular potential for ice at 150\,K. We see that the regularisation has the most noticeable effect on regions with little density -- in the short-range tail of the distribution and in the well between the first and second peaks. Without regularisation, the correction to the potential is clearly noisy in these regions, and it can become very large. The large corrections overshoot the target potential of mean force leading to wild oscillations in the convergence of the IBI, if it converges at all. Even relatively small values of the regularisation parameter are sufficient to remove these instabilities. Furthermore, since each correction to the potential of mean force is now free from statistical errors, which are suppressed by the $\varepsilon\, G_{\rm XY}$ terms in Eq.~\eqref{eq:ibireg}, the final result of the IBI is a potential with smooth forces that is suitable for use in molecular dynamics simulations.

\begin{figure*}[t]
    \centering   \includegraphics[width=.9\textwidth]{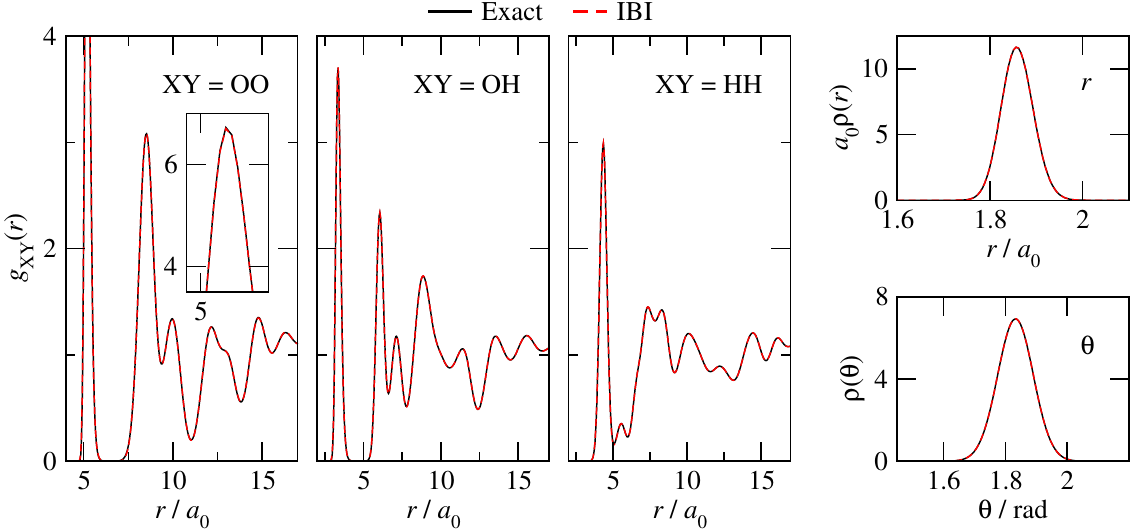}
    \caption{Comparison of exact PIMD and fully converged IBI quasi-centroid distribution functions for hexagonal ice at 150\,K.}
    \label{sifig:ibiconvergence-150K}
\end{figure*}

\subsection{Quasi-centroid molecular dynamics}

Once the IBI has converged, the resulting quasi-centroid potential of mean force $V_{\rm qc}({\bf r})$ can be used to calculate the dipole absorption spectrum from the standard formula
\begin{equation}\label{eq:Iomega}
I(\omega) = {\beta\over 6c\epsilon_0V}\int_{-\infty}^{\infty} e^{-i\omega t}\langle\dot{\bm{\mu}}(0)\cdot\dot{\bm{\mu}}(t)\rangle\,{\rm d}t,
\end{equation}
where the angular brackets denote a canonical ensemble average, the dynamics is purely classical, and $\dot{\bm{\mu}}$ is the time derivative of the dipole moment of the system (here liquid water or ice). This is no more difficult than performing a standard classical molecular dynamics simulation of the spectrum on the potential $V_{\rm cl}({\bf r})$.

\section{Application to liquid water and ice}

\subsection{Computational details}

We have used the IBI algorithm described in Sec.~II to calculate the f-QCMD approximation to the quasi-centroid potential of mean force for q-TIP4P/F water at 300\,K and ice at 150\,K. All simulations used a time step of $0.25$\,fs. The liquid water simulation was perfomed using 216 water molecules in a cubic box with a side length of 35.24 $a_0$, and the ice simulation with 96 molecules in an orthorhombic box with side lengths of 
25.62, 29.58, and 27.89 $a_0$. Following Refs.~\citenum{Tren2019} and \citenum{Tren2022}, the PIMD simulations used $P=32$ beads at 300\,K and $P=64$ beads at 150\,K. We performed a total of 30 regularised IBI iterations at both 300\,K and 150\,K, with $\varepsilon = 1$ and $\varepsilon = 5$ respectively. At both temperatures, the quasi-centroid distribution functions obtained from the IBI were found to be graphically indistinguishable from the target distribution functions provided by the PIMD simulation. The distribution functions for ice are shown in Fig.~\ref{sifig:ibiconvergence-150K} and those for liquid water are shown in Fig.~\ref{sifig:ibiconvergence-300K} of the supplementary material.

At this point we should stress that the agreement between the distribution functions does not guarantee that the potential recovered by the IBI iteration is correct. This has only been shown to be the case in the idealised situation in which  IBI is used to recover a pair potential (to within a constant) from a pair distribution function.\cite{Hend1974} As discussed by Evans,\cite{Evan1990} third and higher order correlation functions are needed to capture higher order interactions. In what we are doing here, the situation is more complicated, because we are not using IBI to recover the potential of mean force itself but rather the (relatively small) difference between it and the underlying classical potential (which does contain higher order interactions). So our IBI solution may not be unique, and we should even be concerned that  using different values of $\varepsilon$ might lead to different results. We have investigated this and found that while varying $\varepsilon$ (within the domain of convergence) does lead to subtle differences in the converged quasi-centroid potential of mean force, these differences do not have any observable effect on the vibrational spectrum.

\begin{figure*}[t]
    \centering  \includegraphics[width=1.0\textwidth]{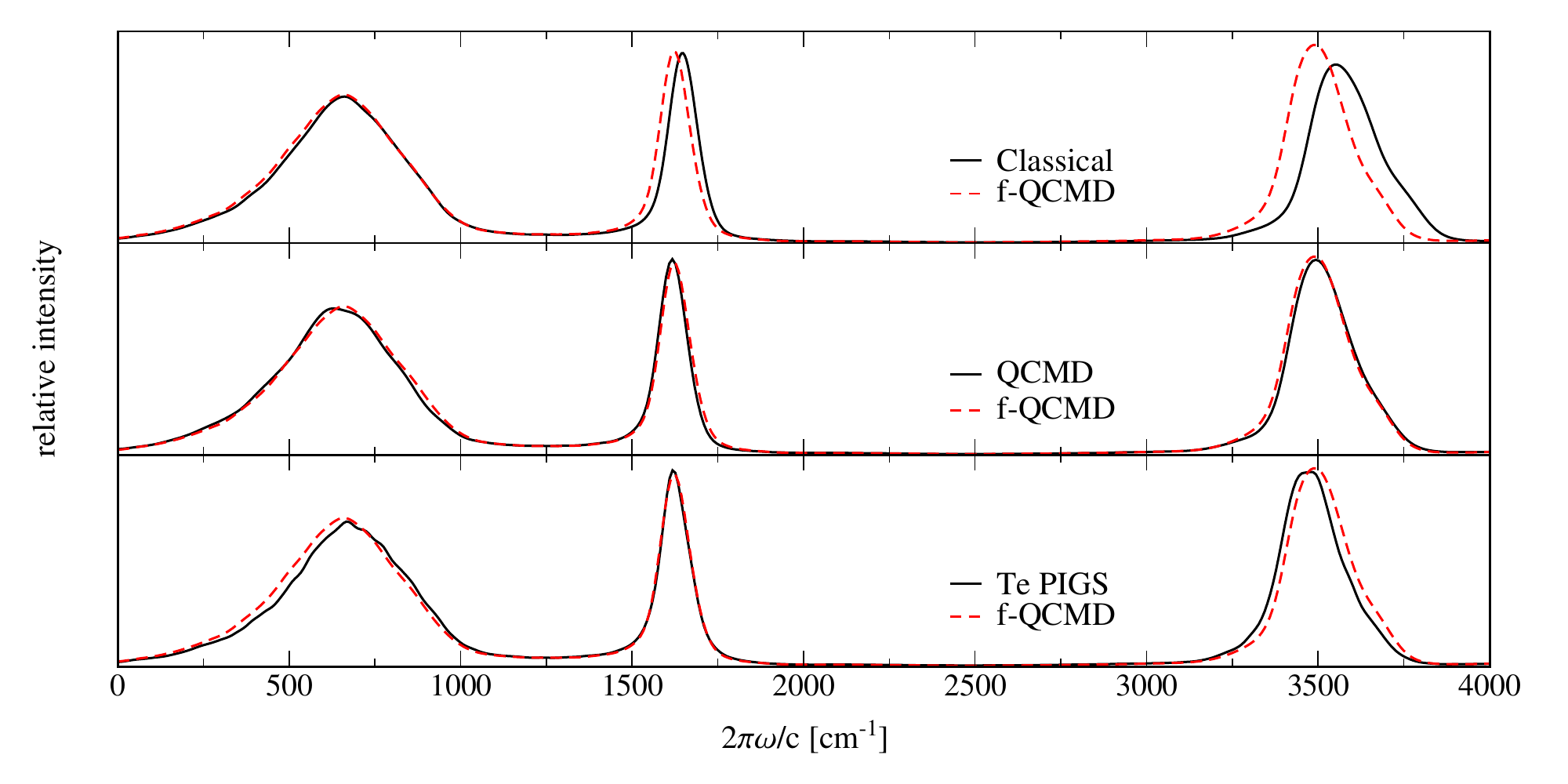}
    \caption{Comparison of classical MD, fast QCMD, adiabatic QCMD, and Te PIGS vibrational spectra of qTIP4P/F water at 300~K.}
    \label{fig:spectrum-300K}
\end{figure*}

\subsection{Liquid water}

Fig.~\ref{fig:spectrum-300K} compares the resulting f-QCMD spectrum for liquid water at 300 K with the spectra obtained from classical molecular dynamics, the adiabatic QCMD method,\cite{Tren2019} and the Te PIGS method.\cite{Musi2022}  (The adiabatic QCMD results were obtained using the new torque estimator described in Ref.~\citenum{Tren2022}, and were provided to us by George Trenins.) All spectra were computed from the dipole-derivative auto-correlation function in Eq.~\eqref{eq:Iomega} damped using a Hann window function with a time constant of 600~fs. Here we are only considering the fundamental region of the spectrum, which has three major peaks: a librational band at 600 cm$^{-1}$, a bending band at 1600 cm$^{-1}$, and an O--H stretching band at 3500 cm$^{-1}$.

The first thing to note is that the agreement between the f-QCMD spectrum and the adiabatic QCMD spectrum is very good, with f-QCMD accurately reproducing both the line shapes and the frequencies of the adiabatic QCMD peaks. This indicates that the separable approximations to the correction potentials in Eqs.~\eqref{eq:Vintra} and \eqref{eq:Vinter} are perfectly adequate for liquid water at 300\,K. The only differences between the two spectra are a very minor red shift in our stretching peak and equally minor blue shifts in our librational and bending bands. 
The much larger differences between the f-QCMD spectrum and the classical spectrum are due to nuclear quantum effects, which result in an anharmonic red shift of around 100 cm$^{-1}$ in the O--H stretching band and a smaller red shift in the bending band. The classical and f-QCMD librational bands are essentially the same because nuclear quantum effects are less significant for low frequency vibrations.

The final panel of Fig.~\ref{fig:spectrum-300K} compares the f-QCMD spectrum with the Te PIGS spectrum obtained using a centroid potential of mean force calculated at 600 K.\cite{Musi2022} The bending bands of the two spectra are the same, but the Te PIGS librational band is slightly blue shifted and the Te PIGS O--H stretching band is slightly red shifted relative to the f-QCMD spectrum. The blue shift in the librational band is most likely a result of incomplete convergence of the auto-correlation function used to calculate the spectrum, as this is the region of the spectrum that converges most slowly. This is corroborated by the slightly uneven line shape of this peak in the Te PIGS calculation. While the slight red shift in the stretching band might also be due to incomplete convergence, it could possibly be a hint of the CMD curvature problem, which is just starting to appear in the centroid potential of mean force at 600~K.\cite{Tren2019} 

\begin{figure*}
    \centering
\includegraphics[width=1.0\textwidth]{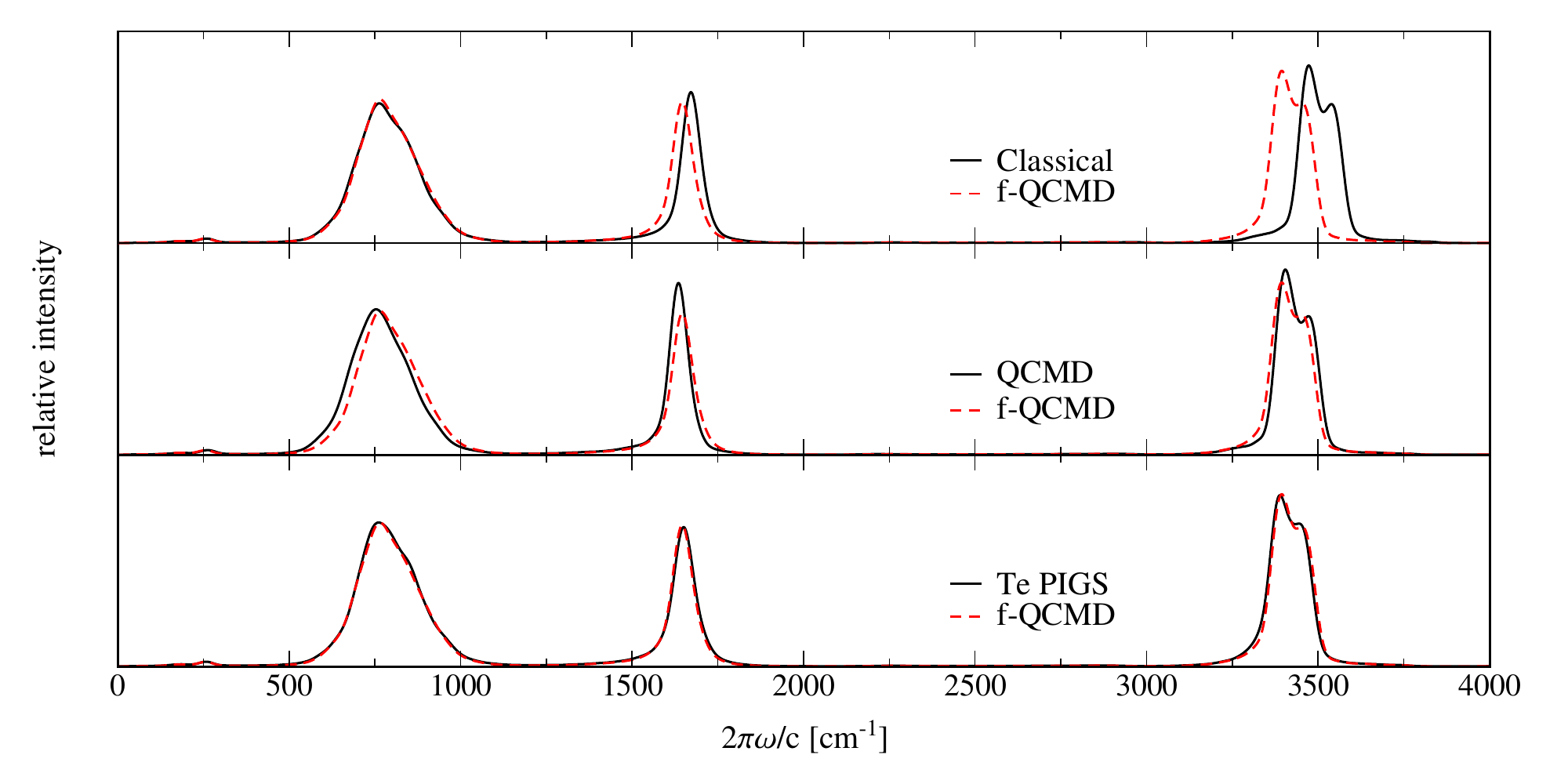}
    \caption{Comparison of classical MD, fast QCMD, adiabatic QCMD, and Te PIGS vibrational spectra of qTIP4P/F ice at 150~K.}
    \label{fig:spectrum-150K}
\end{figure*}

\subsection{Hexagonal ice}

Figure~\ref{fig:spectrum-150K} shows the vibrational spectra for hexagonal ice at 150\,K. These spectra were computed in the same way as the liquid water spectra, but using a Hann window function with a time constant of 800 fs. They differ from the liquid water spectra in that all three of the main bands are narrower, there is more structure in the O--H stretching band, and a tiny inter-molecular O--O stretching band is now apparent at around 250 cm$^{-1}$.

Comparing the QCMD and f-QCMD spectra in Fig.~\ref{fig:spectrum-150K}, it is clear that the differences in ice at 150\,K are more pronounced than they were in water at 300\,K. The f-QCMD bending band is less intense, the librational band is slightly blue shifted, and the stretching band is slightly red shifted relative to QCMD. The most likely explanation for these differences is that the pairwise approximations to the correction potentials in Eqs.~\eqref{eq:Vintra} and~\eqref{eq:Vinter} are less accurate in ice at 150 K than they were in water at 300 K.  Alternatively, the discrepancy could be caused by a lack of convergence with respect to the separation parameters in the adiabatic QCMD algorithm, or by errors introduced by the approximate torque estimator used in QCMD. These possibilities are suggested by  Figs.~\ref{sifig:quasicentroid-300K} and~\ref{sifig:quasicentroid-150K} in the supplementary material, which show that the quasi-centroid distribution functions obtained from the adiabatic QCMD trajectories of Trenins {\em et al.}\cite{Tren2022} do not agree quite so well with our reference PIMD calculations for ice as they do for water. The peak positions in the adiabatic QCMD ice spectrum were found to be converged with respect to the adiabatic separation parameters,\footnote{G. Trenins, private communication (2023).} so this is probably not the root of the  disagreement with f-QCMD.  Regarding the torque estimator, Ref.~\citenum{Tren2022} shows that both QCMD and CMD predict essentially the same red shifts relative to the classical results in the librational and bending bands. As these regions are not expected to be significantly affected by the curvature problem of CMD, this suggests that the errors in the torque estimator are minor. We thus believe that the QCMD peak positions are unlikely to be affected by either of these potential sources of error, although we do not rule out the possibility that the intensities of the QCMD bands may be slightly incorrect.

Interestingly, the agreement between the Te PIGS and f-QCMD spectra in Fig.~\ref{fig:spectrum-150K} is nearly perfect. While this is probably somewhat fortuitous,\footnote{ In particular, the agreement between the stretching bands is likely to be fortuitous. The red shift of Te PIGS relative to QCMD may still be attributed to the onset of the curvature effect at the elevated temperature used to evaluate the PMF, whereas this cannot be the cause of the red shift in f-QCMD.} it is worth speculating about why it might be. The following explanation assumes that the QCMD results are the most accurate in Fig.~\ref{fig:spectrum-150K} and that the Te PIGS results are better converged for ice than they are for water. We have already argued that our pairwise approximation to the correction potential $\Delta V({\bf r})=V_{\rm qc}({\bf r})-V_{\rm cl}({\bf r})$ is justified at high temperatures by the good agreement between f-QCMD and QCMD for liquid water at 300 K. Since the centroid potential of mean force correction used in Te PIGS is calculated at an even higher temperature (600\,K), it too is presumably dominated by pairwise terms. Hence, it is possible that Te PIGS and f-QCMD are both missing the many-body contributions to $\Delta V({\bf r})$ that arise at lower temperatures and lead to the additional red shifts of the librational and bending bands in the QCMD ice spectrum, but for different reasons: Te PIGS because its potential of mean force is computed at a high temperature where these contributions are less significant, and f-QCMD because it inherently assumes that $\Delta V({\bf r})$ can be written in a pairwise form. Note that, since this argument is based on the agreement between f-QCMD and QCMD for water at 300 K (Fig.~\ref{fig:spectrum-300K}), and the agreement between f-QCMD and Te PIGS for ice at 150 K (Fig.~\ref{fig:spectrum-150K}), it would not have been possible to make it without the results of the present calculations.

\subsection{Summary}

In spite of their differences, it is clear from Figs.~\ref{fig:spectrum-300K} and~\ref{fig:spectrum-150K} that all three centroid methods (QCMD, f-QCMD, and Te PIGS) are in better agreement with one another than they are with classical molecular dynamics, and that they provide reasonably consistent predictions of the dominant quantum mechanical effects in the spectra of water and ice (an anharmonic red shift in the intra-molecular stretching band and a smaller red shift in the intra-molecular bending band, with little change in the lineshape of either). This is clear progress compared with the situation a decade or so ago, when the best existing path integral methods were giving markedly different results for these spectra that were plagued by artefacts.  See, for example, Fig.~2 in Ref.~\citenum{Ross2014b} or Figs.~3 and~4 in Ref.~\citenum{Bens2020} for adiabatic CMD\cite{Cao1994d} and thermostatted ring polymer molecular dynamics\cite{Ross2014} spectra that can be compared directly with those in our Figs.~\ref{fig:spectrum-300K} and~\ref{fig:spectrum-150K}. 

One may ask whether these methods not only agree with each other but also whether they agree with the exact result. For a condensed phase system, it is of course not possible to obtain the exact result for comparison. However, for the gas phase water monomer both QCMD and f-QCMD were found to give remarkably good agreement with the quantum reference calculations for the fundamental bands.\cite{Flet2021} This can be understood by noting that in the low temperature limit centroid molecular dynamics gives the exact $0\to1$ transition frequency for an anharmonic oscillator in one dimension,\cite{Rami1998} and that within the coordinate frame used in QCMD the normal modes are only weakly coupled. On going to the condensed phase, one notes that from a single monomer perspective the exact quantum statistics of QCMD means the static effect of the environment is captured exactly. Furthermore the dynamical effect of the environment is captured self-consistently within a classical framework. One can therefore have reasonable confidence that these methods are converging not only to one another but also on an accurate description of the fundamental bands of the vibrational spectrum.

\section{Concluding remarks}

Modern centroid methods are finally reaching a consensus regarding the impact of nuclear quantum effects on the vibrational spectra of water and ice. For the qTIP4P/F water model studied here, they lead to an anharmonic red shift of around 100 cm$^{-1}$ in the intra-molecular O--H stretching band and a smaller red shift in the intra-molecular H--O--H bending band, but very little change in the line shape of either band (see Figs.~\ref{fig:spectrum-300K} and~\ref{fig:spectrum-150K}). Recent Te PIGS calculations demonstrate that this remains largely true for calculations using \emph{ab initio} DFT potential energy surfaces, albeit with larger changes to the intensities and the line shapes of the bands in ice.\cite{Kapi2023}

There will be many situations in which it is reasonable to ignore the relatively small quantum effects we are discussing here and use classical molecular dynamics to simulate the vibrational spectrum. However, there will be other situations where the goal is to understand anharmonic red and blue shifts in vibrational spectra and how they change with temperature and isotopic substitution, which are questions that classical molecular dynamics is incapable of answering.\cite{Litm2020} In these situations, we feel fairly confident on the basis of the present results and previous validation studies for gas phase systems\cite{Tren2019,Hagg2021,Flet2021,Musi2022} (for which exact quantum mechanical benchmark results are available for comparison) that methods like QCMD, f-QCMD, and Te PIGS will provide the right answer.

Given this, we are now in a position to apply these methods to more interesting problems, and we have in fact already completed two such studies using f-QCMD.\cite{Lieb2023} The present quasi-centroid potential of mean force for liquid water has been used to shed light on claims that nuclear quantum effects may broaden vibrational polariton bands,\cite{Lieb2023,Suth2023} and to explore the impact of nuclear quantum effects on the thermal conductivity of liquid water\cite{Suth2023} using the method introduced in Ref.~\onlinecite{Suth2021}. 
For most future applications we would suggest that the more recent Te PIGS method\cite{Musi2022} should be the preferred approach. We have argued in Sec. III.C that Te PIGS is not as accurate as adiabatic QCMD, but it is significantly simpler to implement, and it appears to give comparable accuracy to the pairwise approximation to $\Delta V({\bf r})$ that is made in f-QCMD. It should therefore be perfectly adequate for studying anharmonic effects in many interesting gas and condensed phase systems.

One final comment is that we have focused exclusively on the fundamental bands. Centroid methods are less accurate for overtone and combination bands, most notably failing to capture their quantum mechanical intensity enhancement.\cite{Ple2021,Bens2021} Within the framework of Matsubara dynamics,\cite{Hele2015} this has been shown to be a result of decoupling the centroid mode from the fluctuation modes of the path integral.\cite{Bens2021} An alternative method based on coupling to an effective thermal bath that mimics the so called ``Matsubara heating'' effect has been proposed,\cite{Maug2021} as have post-processing corrections.\cite{Bens2021} Hence, this is clearly an area in which there is still room for improvement on the methods we have considered here.

\section*{Data Availability}
The data to support this study are available in the body of the paper and in the supporting information, code to perform the iterative Boltzmann inversion is available upon request.

\section*{Supporting Information}
Figures showing the convergence of the iterative Boltzmann inversion to the exact distribution functions and the difference between the exact and adiabatic QCMD distribution functions. 

\begin{acknowledgments}
We are grateful to George Trenins for providing the QCMD data for this study and many useful discussions, and to Venkat Kapil for providing the Te PIGS spectra.
Annina Lieberherr was supported by a Berrow Foundation Lord Florey Scholarship and Joseph Lawrence was supported by an ETH Z\"urich Postdoctoral Fellowship. 

\end{acknowledgments}

\section*{References}
\bibliography{fqcmd-condensed}

\begin{thebibliography}{44}%
\makeatletter
\providecommand \@ifxundefined [1]{%
 \@ifx{#1\undefined}
}%
\providecommand \@ifnum [1]{%
 \ifnum #1\expandafter \@firstoftwo
 \else \expandafter \@secondoftwo
 \fi
}%
\providecommand \@ifx [1]{%
 \ifx #1\expandafter \@firstoftwo
 \else \expandafter \@secondoftwo
 \fi
}%
\providecommand \natexlab [1]{#1}%
\providecommand \enquote  [1]{``#1''}%
\providecommand \bibnamefont  [1]{#1}%
\providecommand \bibfnamefont [1]{#1}%
\providecommand \citenamefont [1]{#1}%
\providecommand \href@noop [0]{\@secondoftwo}%
\providecommand \href [0]{\begingroup \@sanitize@url \@href}%
\providecommand \@href[1]{\@@startlink{#1}\@@href}%
\providecommand \@@href[1]{\endgroup#1\@@endlink}%
\providecommand \@sanitize@url [0]{\catcode `\\12\catcode `\$12\catcode
  `\&12\catcode `\#12\catcode `\^12\catcode `\_12\catcode `\%12\relax}%
\providecommand \@@startlink[1]{}%
\providecommand \@@endlink[0]{}%
\providecommand \url  [0]{\begingroup\@sanitize@url \@url }%
\providecommand \@url [1]{\endgroup\@href {#1}{\urlprefix }}%
\providecommand \urlprefix  [0]{URL }%
\providecommand \Eprint [0]{\href }%
\providecommand \doibase [0]{http://dx.doi.org/}%
\providecommand \selectlanguage [0]{\@gobble}%
\providecommand \bibinfo  [0]{\@secondoftwo}%
\providecommand \bibfield  [0]{\@secondoftwo}%
\providecommand \translation [1]{[#1]}%
\providecommand \BibitemOpen [0]{}%
\providecommand \bibitemStop [0]{}%
\providecommand \bibitemNoStop [0]{.\EOS\space}%
\providecommand \EOS [0]{\spacefactor3000\relax}%
\providecommand \BibitemShut  [1]{\csname bibitem#1\endcsname}%
\let\auto@bib@innerbib\@empty
\bibitem [{\citenamefont {Chandler}\ and\ \citenamefont
  {Wolynes}(1981)}]{Chan1981}%
  \BibitemOpen
  \bibfield  {author} {\bibinfo {author} {\bibfnamefont {D.}~\bibnamefont
  {Chandler}}\ and\ \bibinfo {author} {\bibfnamefont {P.~G.}\ \bibnamefont
  {Wolynes}},\ }\bibfield  {title} {\enquote {\bibinfo {title} {Exploiting the
  isomorphism between quantum theory and classical statistical mechanics of
  polyatomic fluids},}\ }\href {\doibase 10.1063/1.441588} {\bibfield
  {journal} {\bibinfo  {journal} {The Journal of Chemical Physics}\ }\textbf
  {\bibinfo {volume} {74}},\ \bibinfo {pages} {4078--4095} (\bibinfo {year}
  {1981})}\BibitemShut {NoStop}%
\bibitem [{\citenamefont {Parrinello}\ and\ \citenamefont
  {Rahman}(1984)}]{Pari1984}%
  \BibitemOpen
  \bibfield  {author} {\bibinfo {author} {\bibfnamefont {M.}~\bibnamefont
  {Parrinello}}\ and\ \bibinfo {author} {\bibfnamefont {A.}~\bibnamefont
  {Rahman}},\ }\bibfield  {title} {\enquote {\bibinfo {title} {Study of an {F}
  center in molten {KC}l},}\ }\href {\doibase 10.1063/1.446740} {\bibfield
  {journal} {\bibinfo  {journal} {The Journal of Chemical Physics}\ }\textbf
  {\bibinfo {volume} {80}},\ \bibinfo {pages} {860--867} (\bibinfo {year}
  {1984})}\BibitemShut {NoStop}%
\bibitem [{\citenamefont {Cao}\ and\ \citenamefont
  {Voth}(1994{\natexlab{a}})}]{Cao1994a}%
  \BibitemOpen
  \bibfield  {author} {\bibinfo {author} {\bibfnamefont {J.}~\bibnamefont
  {Cao}}\ and\ \bibinfo {author} {\bibfnamefont {G.~A.}\ \bibnamefont {Voth}},\
  }\bibfield  {title} {\enquote {\bibinfo {title} {The formulation of quantum
  statistical mechanics based on the {F}eynman path centroid density. {I}.
  {E}quilibrium properties},}\ }\href {\doibase 10.1063/1.467175} {\bibfield
  {journal} {\bibinfo  {journal} {The Journal of Chemical Physics}\ }\textbf
  {\bibinfo {volume} {100}},\ \bibinfo {pages} {5093--5105} (\bibinfo {year}
  {1994}{\natexlab{a}})}\BibitemShut {NoStop}%
\bibitem [{\citenamefont {Cao}\ and\ \citenamefont
  {Voth}(1994{\natexlab{b}})}]{Cao1994b}%
  \BibitemOpen
  \bibfield  {author} {\bibinfo {author} {\bibfnamefont {J.}~\bibnamefont
  {Cao}}\ and\ \bibinfo {author} {\bibfnamefont {G.~A.}\ \bibnamefont {Voth}},\
  }\bibfield  {title} {\enquote {\bibinfo {title} {The formulation of quantum
  statistical mechanics based on the {F}eynman path centroid density. {II}.
  {D}ynamical properties},}\ }\href {\doibase 10.1063/1.467176} {\bibfield
  {journal} {\bibinfo  {journal} {The Journal of Chemical Physics}\ }\textbf
  {\bibinfo {volume} {100}},\ \bibinfo {pages} {5106--5117} (\bibinfo {year}
  {1994}{\natexlab{b}})}\BibitemShut {NoStop}%
\bibitem [{\citenamefont {Cao}\ and\ \citenamefont
  {Voth}(1994{\natexlab{c}})}]{Cao1994c}%
  \BibitemOpen
  \bibfield  {author} {\bibinfo {author} {\bibfnamefont {J.}~\bibnamefont
  {Cao}}\ and\ \bibinfo {author} {\bibfnamefont {G.~A.}\ \bibnamefont {Voth}},\
  }\bibfield  {title} {\enquote {\bibinfo {title} {The formulation of quantum
  statistical mechanics based on the {F}eynman path centroid density. {III}.
  {P}hase space formulation and analysis of centroid molecular dynamics},}\
  }\href {\doibase 10.1063/1.468503} {\bibfield  {journal} {\bibinfo  {journal}
  {The Journal of Chemical Physics}\ }\textbf {\bibinfo {volume} {101}},\
  \bibinfo {pages} {6157--6167} (\bibinfo {year}
  {1994}{\natexlab{c}})}\BibitemShut {NoStop}%
\bibitem [{\citenamefont {Cao}\ and\ \citenamefont
  {Voth}(1994{\natexlab{d}})}]{Cao1994d}%
  \BibitemOpen
  \bibfield  {author} {\bibinfo {author} {\bibfnamefont {J.}~\bibnamefont
  {Cao}}\ and\ \bibinfo {author} {\bibfnamefont {G.~A.}\ \bibnamefont {Voth}},\
  }\bibfield  {title} {\enquote {\bibinfo {title} {The formulation of quantum
  statistical mechanics based on the {F}eynman path centroid density. {IV}.
  {A}lgorithms for centroid molecular dynamics},}\ }\href {\doibase
  10.1063/1.468399} {\bibfield  {journal} {\bibinfo  {journal} {The Journal of
  Chemical Physics}\ }\textbf {\bibinfo {volume} {101}},\ \bibinfo {pages}
  {6168--6183} (\bibinfo {year} {1994}{\natexlab{d}})}\BibitemShut {NoStop}%
\bibitem [{\citenamefont {Feynman}\ and\ \citenamefont
  {Hibbs}(1965)}]{Feyn1965}%
  \BibitemOpen
  \bibfield  {author} {\bibinfo {author} {\bibfnamefont {R.~P.}\ \bibnamefont
  {Feynman}}\ and\ \bibinfo {author} {\bibfnamefont {A.~R.}\ \bibnamefont
  {Hibbs}},\ }\href@noop {} {\emph {\bibinfo {title} {Quantum Mechanics and
  path integrals}}}\ (\bibinfo  {publisher} {McGraw-Hill},\ \bibinfo {address}
  {New York},\ \bibinfo {year} {1965})\BibitemShut {NoStop}%
\bibitem [{\citenamefont {Feynman}\ and\ \citenamefont
  {Kleinert}(1986)}]{Feyn1986}%
  \BibitemOpen
  \bibfield  {author} {\bibinfo {author} {\bibfnamefont {R.~P.}\ \bibnamefont
  {Feynman}}\ and\ \bibinfo {author} {\bibfnamefont {H.}~\bibnamefont
  {Kleinert}},\ }\bibfield  {title} {\enquote {\bibinfo {title} {Effective
  classical partition functions},}\ }\href {\doibase 10.1103/PhysRevA.34.5080}
  {\bibfield  {journal} {\bibinfo  {journal} {Phys. Rev. A}\ }\textbf {\bibinfo
  {volume} {34}},\ \bibinfo {pages} {5080--5084} (\bibinfo {year}
  {1986})}\BibitemShut {NoStop}%
\bibitem [{\citenamefont {Paesani}, \citenamefont {Xantheas},\ and\
  \citenamefont {Voth}(2009)}]{Paes2009}%
  \BibitemOpen
  \bibfield  {author} {\bibinfo {author} {\bibfnamefont {F.}~\bibnamefont
  {Paesani}}, \bibinfo {author} {\bibfnamefont {S.~S.}\ \bibnamefont
  {Xantheas}}, \ and\ \bibinfo {author} {\bibfnamefont {G.~A.}\ \bibnamefont
  {Voth}},\ }\bibfield  {title} {\enquote {\bibinfo {title} {Infrared
  spectroscopy and hydrogen-bond dynamics of liquid water from centroid
  molecular dynamics with an ab initio-based force field},}\ }\href {\doibase
  10.1021/jp907648y} {\bibfield  {journal} {\bibinfo  {journal} {Journal of
  Physical Chemistry B}\ }\textbf {\bibinfo {volume} {113}},\ \bibinfo {pages}
  {13118--13130} (\bibinfo {year} {2009})}\BibitemShut {NoStop}%
\bibitem [{\citenamefont {Paesani}\ and\ \citenamefont
  {Voth}(2010)}]{Paes2010}%
  \BibitemOpen
  \bibfield  {author} {\bibinfo {author} {\bibfnamefont {F.}~\bibnamefont
  {Paesani}}\ and\ \bibinfo {author} {\bibfnamefont {G.~A.}\ \bibnamefont
  {Voth}},\ }\bibfield  {title} {\enquote {\bibinfo {title} {A quantitative
  assessment of the accuracy of centroid molecular dynamics for the calculation
  of the infrared spectrum of liquid water},}\ }\href {\doibase
  10.1063/1.3291212} {\bibfield  {journal} {\bibinfo  {journal} {The Journal of
  Chemical Physics}\ }\textbf {\bibinfo {volume} {132}},\ \bibinfo {pages}
  {014105} (\bibinfo {year} {2010})}\BibitemShut {NoStop}%
\bibitem [{\citenamefont {Ram\'irez}, \citenamefont {L\'opez-Ciudad},\ and\
  \citenamefont {Noya}(1998)}]{Rami1998}%
  \BibitemOpen
  \bibfield  {author} {\bibinfo {author} {\bibfnamefont {R.}~\bibnamefont
  {Ram\'irez}}, \bibinfo {author} {\bibfnamefont {T.}~\bibnamefont
  {L\'opez-Ciudad}}, \ and\ \bibinfo {author} {\bibfnamefont {J.~C.}\
  \bibnamefont {Noya}},\ }\bibfield  {title} {\enquote {\bibinfo {title}
  {Feynman effective classical potential in the schr\"odinger formulation},}\
  }\href {\doibase 10.1103/PhysRevLett.81.3303} {\bibfield  {journal} {\bibinfo
   {journal} {Physical Review Letters}\ }\textbf {\bibinfo {volume} {81}},\
  \bibinfo {pages} {3303--3306} (\bibinfo {year} {1998})}\BibitemShut {NoStop}%
\bibitem [{\citenamefont {Witt}\ \emph {et~al.}(2009)\citenamefont {Witt},
  \citenamefont {Ivanov}, \citenamefont {Shiga}, \citenamefont {Forbert},\ and\
  \citenamefont {Marx}}]{Witt2009}%
  \BibitemOpen
  \bibfield  {author} {\bibinfo {author} {\bibfnamefont {A.}~\bibnamefont
  {Witt}}, \bibinfo {author} {\bibfnamefont {S.~D.}\ \bibnamefont {Ivanov}},
  \bibinfo {author} {\bibfnamefont {M.}~\bibnamefont {Shiga}}, \bibinfo
  {author} {\bibfnamefont {H.}~\bibnamefont {Forbert}}, \ and\ \bibinfo
  {author} {\bibfnamefont {D.}~\bibnamefont {Marx}},\ }\bibfield  {title}
  {\enquote {\bibinfo {title} {On the applicability of centroid and ring
  polymer path integral molecular dynamics for vibrational spectroscopy},}\
  }\href {\doibase 10.1063/1.3125009} {\bibfield  {journal} {\bibinfo
  {journal} {The Journal of Chemical Physics}\ }\textbf {\bibinfo {volume}
  {130}},\ \bibinfo {pages} {194510} (\bibinfo {year} {2009})}\BibitemShut
  {NoStop}%
\bibitem [{\citenamefont {Ivanov}\ \emph {et~al.}(2010)\citenamefont {Ivanov},
  \citenamefont {Witt}, \citenamefont {Shiga},\ and\ \citenamefont
  {Marx}}]{Ivan2010}%
  \BibitemOpen
  \bibfield  {author} {\bibinfo {author} {\bibfnamefont {S.~D.}\ \bibnamefont
  {Ivanov}}, \bibinfo {author} {\bibfnamefont {A.}~\bibnamefont {Witt}},
  \bibinfo {author} {\bibfnamefont {M.}~\bibnamefont {Shiga}}, \ and\ \bibinfo
  {author} {\bibfnamefont {D.}~\bibnamefont {Marx}},\ }\bibfield  {title}
  {\enquote {\bibinfo {title} {On artificial frequency shifts in infrared
  spectra obtained from centroid molecular dynamics: Quantum liquid water},}\
  }\href {\doibase 10.1063/1.3290958} {\bibfield  {journal} {\bibinfo
  {journal} {The Journal of Chemical Physics}\ }\textbf {\bibinfo {volume}
  {132}},\ \bibinfo {pages} {031101} (\bibinfo {year} {2010})}\BibitemShut
  {NoStop}%
\bibitem [{\citenamefont {Benson}, \citenamefont {Trenins},\ and\ \citenamefont
  {Althorpe}(2020)}]{Bens2020}%
  \BibitemOpen
  \bibfield  {author} {\bibinfo {author} {\bibfnamefont {R.~L.}\ \bibnamefont
  {Benson}}, \bibinfo {author} {\bibfnamefont {G.}~\bibnamefont {Trenins}}, \
  and\ \bibinfo {author} {\bibfnamefont {S.~C.}\ \bibnamefont {Althorpe}},\
  }\bibfield  {title} {\enquote {\bibinfo {title} {Which quantum
  statistics{\textendash}classical dynamics method is best for water?}}\ }\href
  {\doibase 10.1039/c9fd00077a} {\bibfield  {journal} {\bibinfo  {journal}
  {Faraday Discussions}\ }\textbf {\bibinfo {volume} {221}},\ \bibinfo {pages}
  {350--366} (\bibinfo {year} {2020})}\BibitemShut {NoStop}%
\bibitem [{\citenamefont {Trenins}, \citenamefont {Willatt},\ and\
  \citenamefont {Althorpe}(2019)}]{Tren2019}%
  \BibitemOpen
  \bibfield  {author} {\bibinfo {author} {\bibfnamefont {G.}~\bibnamefont
  {Trenins}}, \bibinfo {author} {\bibfnamefont {M.~J.}\ \bibnamefont
  {Willatt}}, \ and\ \bibinfo {author} {\bibfnamefont {S.~C.}\ \bibnamefont
  {Althorpe}},\ }\bibfield  {title} {\enquote {\bibinfo {title} {Path-integral
  dynamics of water using curvilinear centroids},}\ }\href {\doibase
  10.1063/1.5100587} {\bibfield  {journal} {\bibinfo  {journal} {The Journal of
  Chemical Physics}\ }\textbf {\bibinfo {volume} {151}},\ \bibinfo {pages}
  {054109} (\bibinfo {year} {2019})}\BibitemShut {NoStop}%
\bibitem [{\citenamefont {Haggard}\ \emph {et~al.}(2021)\citenamefont
  {Haggard}, \citenamefont {Sadhasivam}, \citenamefont {Trenins},\ and\
  \citenamefont {Althorpe}}]{Hagg2021}%
  \BibitemOpen
  \bibfield  {author} {\bibinfo {author} {\bibfnamefont {C.}~\bibnamefont
  {Haggard}}, \bibinfo {author} {\bibfnamefont {V.~G.}\ \bibnamefont
  {Sadhasivam}}, \bibinfo {author} {\bibfnamefont {G.}~\bibnamefont {Trenins}},
  \ and\ \bibinfo {author} {\bibfnamefont {S.~C.}\ \bibnamefont {Althorpe}},\
  }\bibfield  {title} {\enquote {\bibinfo {title} {Testing the quasicentroid
  molecular dynamics method on gas-phase ammonia},}\ }\href {\doibase
  10.1063/5.0068250} {\bibfield  {journal} {\bibinfo  {journal} {The Journal of
  Chemical Physics}\ }\textbf {\bibinfo {volume} {155}},\ \bibinfo {pages}
  {174120} (\bibinfo {year} {2021})}\BibitemShut {NoStop}%
\bibitem [{\citenamefont {Trenins}, \citenamefont {Haggard},\ and\
  \citenamefont {Althorpe}(2022)}]{Tren2022}%
  \BibitemOpen
  \bibfield  {author} {\bibinfo {author} {\bibfnamefont {G.}~\bibnamefont
  {Trenins}}, \bibinfo {author} {\bibfnamefont {C.}~\bibnamefont {Haggard}}, \
  and\ \bibinfo {author} {\bibfnamefont {S.~C.}\ \bibnamefont {Althorpe}},\
  }\bibfield  {title} {\enquote {\bibinfo {title} {{Improved torque estimator
  for condensed-phase quasicentroid molecular dynamics}},}\ }\href {\doibase
  10.1063/5.0129482} {\bibfield  {journal} {\bibinfo  {journal} {The Journal of
  Chemical Physics}\ }\textbf {\bibinfo {volume} {157}},\ \bibinfo {pages}
  {174108} (\bibinfo {year} {2022})}\BibitemShut {NoStop}%
\bibitem [{\citenamefont {Fletcher}\ \emph {et~al.}(2021)\citenamefont
  {Fletcher}, \citenamefont {Zhu}, \citenamefont {Lawrence},\ and\
  \citenamefont {Manolopoulos}}]{Flet2021}%
  \BibitemOpen
  \bibfield  {author} {\bibinfo {author} {\bibfnamefont {T.}~\bibnamefont
  {Fletcher}}, \bibinfo {author} {\bibfnamefont {A.}~\bibnamefont {Zhu}},
  \bibinfo {author} {\bibfnamefont {J.~E.}\ \bibnamefont {Lawrence}}, \ and\
  \bibinfo {author} {\bibfnamefont {D.~E.}\ \bibnamefont {Manolopoulos}},\
  }\bibfield  {title} {\enquote {\bibinfo {title} {Fast quasi-centroid
  molecular dynamics},}\ }\href {\doibase 10.1063/5.0076704} {\bibfield
  {journal} {\bibinfo  {journal} {The Journal of Chemical Physics}\ }\textbf
  {\bibinfo {volume} {155}},\ \bibinfo {pages} {231101} (\bibinfo {year}
  {2021})}\BibitemShut {NoStop}%
\bibitem [{\citenamefont {Hone}, \citenamefont {Izvekov},\ and\ \citenamefont
  {Voth}(2005)}]{Hone2005}%
  \BibitemOpen
  \bibfield  {author} {\bibinfo {author} {\bibfnamefont {T.~D.}\ \bibnamefont
  {Hone}}, \bibinfo {author} {\bibfnamefont {S.}~\bibnamefont {Izvekov}}, \
  and\ \bibinfo {author} {\bibfnamefont {G.~A.}\ \bibnamefont {Voth}},\
  }\bibfield  {title} {\enquote {\bibinfo {title} {Fast centroid molecular
  dynamics: {A} force-matching approach for the predetermination of the
  effective centroid forces},}\ }\href {\doibase 10.1063/1.1836731} {\bibfield
  {journal} {\bibinfo  {journal} {The Journal of Chemical Physics}\ }\textbf
  {\bibinfo {volume} {122}},\ \bibinfo {pages} {054105} (\bibinfo {year}
  {2005})}\BibitemShut {NoStop}%
\bibitem [{\citenamefont {Soper}(1996)}]{Sope1996}%
  \BibitemOpen
  \bibfield  {author} {\bibinfo {author} {\bibfnamefont {A.~K.}\ \bibnamefont
  {Soper}},\ }\bibfield  {title} {\enquote {\bibinfo {title} {Empirical
  potential {M}onte {C}arlo simulation of fluid structure},}\ }\href {\doibase
  10.1016/0301-0104(95)00357-6} {\bibfield  {journal} {\bibinfo  {journal}
  {Chemical Physics}\ }\textbf {\bibinfo {volume} {202}},\ \bibinfo {pages}
  {295--306} (\bibinfo {year} {1996})}\BibitemShut {NoStop}%
\bibitem [{\citenamefont {Reith}, \citenamefont {P\"utz},\ and\ \citenamefont
  {M\"uller-Plathe}(2003)}]{Reit2003}%
  \BibitemOpen
  \bibfield  {author} {\bibinfo {author} {\bibfnamefont {D.}~\bibnamefont
  {Reith}}, \bibinfo {author} {\bibfnamefont {M.}~\bibnamefont {P\"utz}}, \
  and\ \bibinfo {author} {\bibfnamefont {F.}~\bibnamefont {M\"uller-Plathe}},\
  }\bibfield  {title} {\enquote {\bibinfo {title} {Deriving effective mesoscale
  potentials from atomistic simulations},}\ }\href {\doibase 10.1002/jcc.10307}
  {\bibfield  {journal} {\bibinfo  {journal} {Journal of Computational
  Chemistry}\ }\textbf {\bibinfo {volume} {24}},\ \bibinfo {pages} {1624--1636}
  (\bibinfo {year} {2003})}\BibitemShut {NoStop}%
\bibitem [{\citenamefont {Musil}\ \emph {et~al.}(2022)\citenamefont {Musil},
  \citenamefont {Zaporozhets}, \citenamefont {No{\'{e}}}, \citenamefont
  {Clementi},\ and\ \citenamefont {Kapil}}]{Musi2022}%
  \BibitemOpen
  \bibfield  {author} {\bibinfo {author} {\bibfnamefont {F.}~\bibnamefont
  {Musil}}, \bibinfo {author} {\bibfnamefont {I.}~\bibnamefont {Zaporozhets}},
  \bibinfo {author} {\bibfnamefont {F.}~\bibnamefont {No{\'{e}}}}, \bibinfo
  {author} {\bibfnamefont {C.}~\bibnamefont {Clementi}}, \ and\ \bibinfo
  {author} {\bibfnamefont {V.}~\bibnamefont {Kapil}},\ }\bibfield  {title}
  {\enquote {\bibinfo {title} {Quantum dynamics using path integral
  coarse-graining},}\ }\href {\doibase 10.1063/5.0120386} {\bibfield  {journal}
  {\bibinfo  {journal} {The Journal of Chemical Physics}\ }\textbf {\bibinfo
  {volume} {157}},\ \bibinfo {pages} {181102} (\bibinfo {year}
  {2022})}\BibitemShut {NoStop}%
\bibitem [{\citenamefont {Krasnoshchekov}, \citenamefont {Isayeva},\ and\
  \citenamefont {Stepanov}(2014)}]{Kras2014}%
  \BibitemOpen
  \bibfield  {author} {\bibinfo {author} {\bibfnamefont {S.~V.}\ \bibnamefont
  {Krasnoshchekov}}, \bibinfo {author} {\bibfnamefont {E.~V.}\ \bibnamefont
  {Isayeva}}, \ and\ \bibinfo {author} {\bibfnamefont {N.~F.}\ \bibnamefont
  {Stepanov}},\ }\bibfield  {title} {\enquote {\bibinfo {title} {{Determination
  of the {E}ckart molecule-fixed frame by use of the apparatus of quaternion
  algebra}},}\ }\href {\doibase 10.1063/1.4870936} {\bibfield  {journal}
  {\bibinfo  {journal} {The Journal of Chemical Physics}\ }\textbf {\bibinfo
  {volume} {140}},\ \bibinfo {pages} {154104} (\bibinfo {year}
  {2014})}\BibitemShut {NoStop}%
\bibitem [{\citenamefont {Lyubartsev}\ and\ \citenamefont
  {Laaksonen}(1995)}]{Lyub1995}%
  \BibitemOpen
  \bibfield  {author} {\bibinfo {author} {\bibfnamefont {A.~P.}\ \bibnamefont
  {Lyubartsev}}\ and\ \bibinfo {author} {\bibfnamefont {A.}~\bibnamefont
  {Laaksonen}},\ }\bibfield  {title} {\enquote {\bibinfo {title} {Calculation
  of effective interaction potentials from radial distribution functions: {A}
  reverse {M}onte {C}arlo approach},}\ }\href {\doibase
  10.1103/physreve.52.3730} {\bibfield  {journal} {\bibinfo  {journal}
  {Physical Review E}\ }\textbf {\bibinfo {volume} {52}},\ \bibinfo {pages}
  {3730--3737} (\bibinfo {year} {1995})}\BibitemShut {NoStop}%
\bibitem [{\citenamefont {McGreevy}\ and\ \citenamefont
  {Pusztai}(1988)}]{McGr1988}%
  \BibitemOpen
  \bibfield  {author} {\bibinfo {author} {\bibfnamefont {R.~L.}\ \bibnamefont
  {McGreevy}}\ and\ \bibinfo {author} {\bibfnamefont {L.}~\bibnamefont
  {Pusztai}},\ }\bibfield  {title} {\enquote {\bibinfo {title} {Reverse {M}onte
  {C}arlo simulation: {A} new technique for the determination of disordered
  structures},}\ }\href {\doibase 10.1080/08927028808080958} {\bibfield
  {journal} {\bibinfo  {journal} {Molecular Simulation}\ }\textbf {\bibinfo
  {volume} {1}},\ \bibinfo {pages} {359--367} (\bibinfo {year}
  {1988})}\BibitemShut {NoStop}%
\bibitem [{\citenamefont {Coles}\ \emph {et~al.}(2021)\citenamefont {Coles},
  \citenamefont {Mangaud}, \citenamefont {Frenkel},\ and\ \citenamefont
  {Rotenberg}}]{Cole2021}%
  \BibitemOpen
  \bibfield  {author} {\bibinfo {author} {\bibfnamefont {S.~W.}\ \bibnamefont
  {Coles}}, \bibinfo {author} {\bibfnamefont {E.}~\bibnamefont {Mangaud}},
  \bibinfo {author} {\bibfnamefont {D.}~\bibnamefont {Frenkel}}, \ and\
  \bibinfo {author} {\bibfnamefont {B.}~\bibnamefont {Rotenberg}},\ }\bibfield
  {title} {\enquote {\bibinfo {title} {Reduced variance analysis of molecular
  dynamics simulations by linear combination of estimators},}\ }\href {\doibase
  10.1063/5.0053737} {\bibfield  {journal} {\bibinfo  {journal} {The Journal of
  Chemical Physics}\ }\textbf {\bibinfo {volume} {154}},\ \bibinfo {pages}
  {191101} (\bibinfo {year} {2021})}\BibitemShut {NoStop}%
\bibitem [{\citenamefont {Rotenberg}(2020)}]{Rote2020}%
  \BibitemOpen
  \bibfield  {author} {\bibinfo {author} {\bibfnamefont {B.}~\bibnamefont
  {Rotenberg}},\ }\bibfield  {title} {\enquote {\bibinfo {title} {Use the
  force! reduced variance estimators for densities, radial distribution
  functions, and local mobilities in molecular simulations},}\ }\href {\doibase
  10.1063/5.0029113} {\bibfield  {journal} {\bibinfo  {journal} {The Journal of
  Chemical Physics}\ }\textbf {\bibinfo {volume} {153}},\ \bibinfo {pages}
  {150902} (\bibinfo {year} {2020})}\BibitemShut {NoStop}%
\bibitem [{\citenamefont {Habershon}, \citenamefont {Markland},\ and\
  \citenamefont {Manolopoulos}(2009)}]{Habe2009}%
  \BibitemOpen
  \bibfield  {author} {\bibinfo {author} {\bibfnamefont {S.}~\bibnamefont
  {Habershon}}, \bibinfo {author} {\bibfnamefont {T.~E.}\ \bibnamefont
  {Markland}}, \ and\ \bibinfo {author} {\bibfnamefont {D.~E.}\ \bibnamefont
  {Manolopoulos}},\ }\bibfield  {title} {\enquote {\bibinfo {title} {Competing
  quantum effects in the dynamics of a flexible water model},}\ }\href
  {\doibase 10.1063/1.3167790} {\bibfield  {journal} {\bibinfo  {journal} {The
  Journal of Chemical Physics}\ }\textbf {\bibinfo {volume} {131}},\ \bibinfo
  {pages} {024501} (\bibinfo {year} {2009})}\BibitemShut {NoStop}%
\bibitem [{\citenamefont {Hanke}(2017)}]{Hank2017}%
  \BibitemOpen
  \bibfield  {author} {\bibinfo {author} {\bibfnamefont {M.}~\bibnamefont
  {Hanke}},\ }\bibfield  {title} {\enquote {\bibinfo {title} {Well-posedness of
  the iterative {B}oltzmann inversion},}\ }\href {\doibase
  10.1007/s10955-017-1944-2} {\bibfield  {journal} {\bibinfo  {journal}
  {Journal of Statistical Physics}\ }\textbf {\bibinfo {volume} {170}},\
  \bibinfo {pages} {536--553} (\bibinfo {year} {2017})}\BibitemShut {NoStop}%
\bibitem [{\citenamefont {Henderson}(1974)}]{Hend1974}%
  \BibitemOpen
  \bibfield  {author} {\bibinfo {author} {\bibfnamefont {R.~L.}\ \bibnamefont
  {Henderson}},\ }\bibfield  {title} {\enquote {\bibinfo {title} {A uniqueness
  theorem for fluid pair correlation functions},}\ }\href {\doibase
  10.1016/0375-9601(74)90847-0} {\bibfield  {journal} {\bibinfo  {journal}
  {Physics Letters A}\ }\textbf {\bibinfo {volume} {49}},\ \bibinfo {pages}
  {197--198} (\bibinfo {year} {1974})}\BibitemShut {NoStop}%
\bibitem [{\citenamefont {Evans}(1990)}]{Evan1990}%
  \BibitemOpen
  \bibfield  {author} {\bibinfo {author} {\bibfnamefont {R.}~\bibnamefont
  {Evans}},\ }\bibfield  {title} {\enquote {\bibinfo {title} {Comment on
  reverse {M}onte {C}arlo simulation},}\ }\href {\doibase
  10.1080/08927029008022403} {\bibfield  {journal} {\bibinfo  {journal}
  {Molecular Simulation}\ }\textbf {\bibinfo {volume} {4}},\ \bibinfo {pages}
  {409--411} (\bibinfo {year} {1990})}\BibitemShut {NoStop}%
\bibitem [{Note1()}]{Note1}%
  \BibitemOpen
  \bibinfo {note} {G. Trenins, private communication (2023).}\BibitemShut
  {Stop}%
\bibitem [{Note2()}]{Note2}%
  \BibitemOpen
  \bibinfo {note} {In particular, the agreement between the stretching bands is
  likely to be fortuitous. The red shift of Te PIGS relative to QCMD may still
  be attributed to the onset of the curvature effect at the elevated
  temperature used to evaluate the PMF, whereas this cannot be the cause of the
  red shift in f-QCMD.}\BibitemShut {Stop}%
\bibitem [{\citenamefont {Rossi}\ \emph {et~al.}(2014)\citenamefont {Rossi},
  \citenamefont {Liu}, \citenamefont {Paesani}, \citenamefont {Bowman},\ and\
  \citenamefont {Ceriotti}}]{Ross2014b}%
  \BibitemOpen
  \bibfield  {author} {\bibinfo {author} {\bibfnamefont {M.}~\bibnamefont
  {Rossi}}, \bibinfo {author} {\bibfnamefont {H.}~\bibnamefont {Liu}}, \bibinfo
  {author} {\bibfnamefont {F.}~\bibnamefont {Paesani}}, \bibinfo {author}
  {\bibfnamefont {J.}~\bibnamefont {Bowman}}, \ and\ \bibinfo {author}
  {\bibfnamefont {M.}~\bibnamefont {Ceriotti}},\ }\bibfield  {title} {\enquote
  {\bibinfo {title} {Communication: On the consistency of approximate quantum
  dynamics simulation methods for vibrational spectra in the condensed
  phase},}\ }\href {\doibase 10.1063/1.4901214} {\bibfield  {journal} {\bibinfo
   {journal} {The Journal of Chemical Physics}\ }\textbf {\bibinfo {volume}
  {141}},\ \bibinfo {pages} {181101} (\bibinfo {year} {2014})}\BibitemShut
  {NoStop}%
\bibitem [{\citenamefont {Rossi}, \citenamefont {Ceriotti},\ and\ \citenamefont
  {Manolopoulos}(2014)}]{Ross2014}%
  \BibitemOpen
  \bibfield  {author} {\bibinfo {author} {\bibfnamefont {M.}~\bibnamefont
  {Rossi}}, \bibinfo {author} {\bibfnamefont {M.}~\bibnamefont {Ceriotti}}, \
  and\ \bibinfo {author} {\bibfnamefont {D.~E.}\ \bibnamefont {Manolopoulos}},\
  }\bibfield  {title} {\enquote {\bibinfo {title} {How to remove the spurious
  resonances from ring polymer molecular dynamics},}\ }\href {\doibase
  10.1063/1.4883861} {\bibfield  {journal} {\bibinfo  {journal} {The Journal of
  Chemical Physics}\ }\textbf {\bibinfo {volume} {140}},\ \bibinfo {pages}
  {234116} (\bibinfo {year} {2014})}\BibitemShut {NoStop}%
\bibitem [{\citenamefont {Kapil}\ \emph {et~al.}(2023)\citenamefont {Kapil},
  \citenamefont {Kovacs}, \citenamefont {Cs\'anyi},\ and\ \citenamefont
  {Michaelides}}]{Kapi2023}%
  \BibitemOpen
  \bibfield  {author} {\bibinfo {author} {\bibfnamefont {V.}~\bibnamefont
  {Kapil}}, \bibinfo {author} {\bibfnamefont {D.~P.}\ \bibnamefont {Kovacs}},
  \bibinfo {author} {\bibfnamefont {G.}~\bibnamefont {Cs\'anyi}}, \ and\
  \bibinfo {author} {\bibfnamefont {A.}~\bibnamefont {Michaelides}},\
  }\bibfield  {title} {\enquote {\bibinfo {title} {First-principles
  spectroscopy of aqueous interfaces using machine-learned electronic and
  quantum nuclear effects},}\ }\href {\doibase 10.1039/D3FD00113J} {\bibfield
  {journal} {\bibinfo  {journal} {Faraday Discussions}\ }\textbf {\bibinfo
  {volume} {Accepted Manuscript}},\ \bibinfo {pages} {10.1039/D3FD00113J}
  (\bibinfo {year} {2023})}\BibitemShut {NoStop}%
\bibitem [{\citenamefont {Litman}, \citenamefont {Behler},\ and\ \citenamefont
  {Rossi}(2020)}]{Litm2020}%
  \BibitemOpen
  \bibfield  {author} {\bibinfo {author} {\bibfnamefont {Y.}~\bibnamefont
  {Litman}}, \bibinfo {author} {\bibfnamefont {J.}~\bibnamefont {Behler}}, \
  and\ \bibinfo {author} {\bibfnamefont {M.}~\bibnamefont {Rossi}},\ }\bibfield
   {title} {\enquote {\bibinfo {title} {Temperature dependence of the
  vibrational spectrum of porphycene: a qualitative failure of classical-nuclei
  molecular dynamics},}\ }\href {\doibase 10.1039/c9fd00056a} {\bibfield
  {journal} {\bibinfo  {journal} {Faraday Discussions}\ }\textbf {\bibinfo
  {volume} {221}},\ \bibinfo {pages} {526--546} (\bibinfo {year}
  {2020})}\BibitemShut {NoStop}%
\bibitem [{\citenamefont {Lieberherr}\ \emph {et~al.}(2023)\citenamefont
  {Lieberherr}, \citenamefont {Furniss}, \citenamefont {Lawrence},\ and\
  \citenamefont {Manolopoulos}}]{Lieb2023}%
  \BibitemOpen
  \bibfield  {author} {\bibinfo {author} {\bibfnamefont {A.~Z.}\ \bibnamefont
  {Lieberherr}}, \bibinfo {author} {\bibfnamefont {S.~T.~E.}\ \bibnamefont
  {Furniss}}, \bibinfo {author} {\bibfnamefont {J.~E.}\ \bibnamefont
  {Lawrence}}, \ and\ \bibinfo {author} {\bibfnamefont {D.~E.}\ \bibnamefont
  {Manolopoulos}},\ }\bibfield  {title} {\enquote {\bibinfo {title}
  {Vibrational strong coupling in liquid water from cavity molecular
  dynamics},}\ }\href {\doibase 10.1063/1.0156808} {\bibfield  {journal}
  {\bibinfo  {journal} {The Journal of Chemical Physics}\ }\textbf {\bibinfo
  {volume} {158}},\ \bibinfo {pages} {234106} (\bibinfo {year}
  {2023})}\BibitemShut {NoStop}%
\bibitem [{\citenamefont {Sutherland}(2023)}]{Suth2023}%
  \BibitemOpen
  \bibfield  {author} {\bibinfo {author} {\bibfnamefont {B.~J.}\ \bibnamefont
  {Sutherland}},\ }\href@noop {} {\emph {\bibinfo {title} {Inclusion of nuclear
  quantum effects in thermal conductivity prediction from path-integral
  techniques}}}\ (\bibinfo  {publisher} {D.Phil. thesis, Oxford University},\
  \bibinfo {year} {2023})\BibitemShut {NoStop}%
\bibitem [{\citenamefont {Sutherland}, \citenamefont {Moore},\ and\
  \citenamefont {Manolopoulos}(2021)}]{Suth2021}%
  \BibitemOpen
  \bibfield  {author} {\bibinfo {author} {\bibfnamefont {B.~J.}\ \bibnamefont
  {Sutherland}}, \bibinfo {author} {\bibfnamefont {W.~H.~D.}\ \bibnamefont
  {Moore}}, \ and\ \bibinfo {author} {\bibfnamefont {D.~E.}\ \bibnamefont
  {Manolopoulos}},\ }\bibfield  {title} {\enquote {\bibinfo {title} {Nuclear
  quantum effects in thermal conductivity from centroid molecular dynamics},}\
  }\href {\doibase 10.1063/1.0051663} {\bibfield  {journal} {\bibinfo
  {journal} {The Journal of Chemical Physics}\ }\textbf {\bibinfo {volume}
  {154}},\ \bibinfo {pages} {174104} (\bibinfo {year} {2021})}\BibitemShut
  {NoStop}%
\bibitem [{\citenamefont {Pl\'e}\ \emph {et~al.}(2021)\citenamefont {Pl\'e},
  \citenamefont {Huppert}, \citenamefont {Finocchi}, \citenamefont {Depondt},\
  and\ \citenamefont {Bonella}}]{Ple2021}%
  \BibitemOpen
  \bibfield  {author} {\bibinfo {author} {\bibfnamefont {T.}~\bibnamefont
  {Pl\'e}}, \bibinfo {author} {\bibfnamefont {S.}~\bibnamefont {Huppert}},
  \bibinfo {author} {\bibfnamefont {F.}~\bibnamefont {Finocchi}}, \bibinfo
  {author} {\bibfnamefont {P.}~\bibnamefont {Depondt}}, \ and\ \bibinfo
  {author} {\bibfnamefont {S.}~\bibnamefont {Bonella}},\ }\bibfield  {title}
  {\enquote {\bibinfo {title} {Anharmonic spectral features via
  trajectory-based quantum dynamics: A perturbative analysis of the interplay
  between dynamics and sampling},}\ }\href {\doibase 10.1063/5.0056824}
  {\bibfield  {journal} {\bibinfo  {journal} {The Journal of Chemical Physics}\
  }\textbf {\bibinfo {volume} {155}},\ \bibinfo {pages} {104108} (\bibinfo
  {year} {2021})}\BibitemShut {NoStop}%
\bibitem [{\citenamefont {Benson}\ and\ \citenamefont
  {Althorpe}(2021)}]{Bens2021}%
  \BibitemOpen
  \bibfield  {author} {\bibinfo {author} {\bibfnamefont {R.~L.}\ \bibnamefont
  {Benson}}\ and\ \bibinfo {author} {\bibfnamefont {S.~C.}\ \bibnamefont
  {Althorpe}},\ }\bibfield  {title} {\enquote {\bibinfo {title} {On the
  ``matsubara heating" of overtone intensities and fermi splittings},}\ }\href
  {\doibase 10.1063/5.0056829} {\bibfield  {journal} {\bibinfo  {journal} {The
  Journal of Chemical Physics}\ }\textbf {\bibinfo {volume} {155}},\ \bibinfo
  {pages} {104107} (\bibinfo {year} {2021})}\BibitemShut {NoStop}%
\bibitem [{\citenamefont {Hele}\ \emph {et~al.}(2015)\citenamefont {Hele},
  \citenamefont {Willatt}, \citenamefont {Muolo},\ and\ \citenamefont
  {Althorpe}}]{Hele2015}%
  \BibitemOpen
  \bibfield  {author} {\bibinfo {author} {\bibfnamefont {T.~J.~H.}\
  \bibnamefont {Hele}}, \bibinfo {author} {\bibfnamefont {M.~J.}\ \bibnamefont
  {Willatt}}, \bibinfo {author} {\bibfnamefont {A.}~\bibnamefont {Muolo}}, \
  and\ \bibinfo {author} {\bibfnamefont {S.~C.}\ \bibnamefont {Althorpe}},\
  }\bibfield  {title} {\enquote {\bibinfo {title} {Boltzmann-conserving
  classical dynamics in quantum time-correlation functions: ``matsubara
  dynamics"},}\ }\href {\doibase 10.1063/1.4916311} {\bibfield  {journal}
  {\bibinfo  {journal} {The Journal of Chemical Physics}\ }\textbf {\bibinfo
  {volume} {142}},\ \bibinfo {pages} {134103} (\bibinfo {year}
  {2015})}\BibitemShut {NoStop}%
\bibitem [{\citenamefont {Mauger}\ \emph {et~al.}(2021)\citenamefont {Mauger},
  \citenamefont {Pl\'e}, \citenamefont {Lagard\`ere}, \citenamefont {Bonella},
  \citenamefont {Mangaud}, \citenamefont {Piquemal},\ and\ \citenamefont
  {Huppert}}]{Maug2021}%
  \BibitemOpen
  \bibfield  {author} {\bibinfo {author} {\bibfnamefont {N.}~\bibnamefont
  {Mauger}}, \bibinfo {author} {\bibfnamefont {T.}~\bibnamefont {Pl\'e}},
  \bibinfo {author} {\bibfnamefont {L.}~\bibnamefont {Lagard\`ere}}, \bibinfo
  {author} {\bibfnamefont {S.}~\bibnamefont {Bonella}}, \bibinfo {author}
  {\bibfnamefont {E.}~\bibnamefont {Mangaud}}, \bibinfo {author} {\bibfnamefont
  {J.-P.}\ \bibnamefont {Piquemal}}, \ and\ \bibinfo {author} {\bibfnamefont
  {S.}~\bibnamefont {Huppert}},\ }\bibfield  {title} {\enquote {\bibinfo
  {title} {Nuclear quantum effects in liquid water at near classical
  computational cost using the adaptive quantum thermal bath},}\ }\href
  {\doibase 10.1021/acs.jpclett.1c01722} {\bibfield  {journal} {\bibinfo
  {journal} {Journal of Physical Chemistry Letters}\ }\textbf {\bibinfo
  {volume} {12}},\ \bibinfo {pages} {8285--8291} (\bibinfo {year}
  {2021})}\BibitemShut {NoStop}%
\end{thebibliography}%

\begin{figure*}
    \centering
\includegraphics{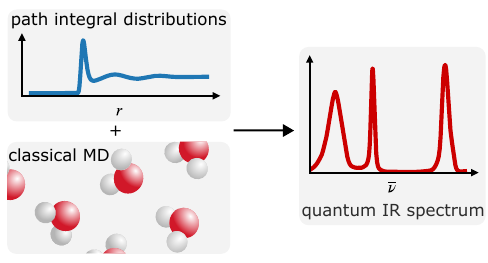}
    \caption*{TOC Graphic}
\end{figure*}
\end{document}


\maketitle


\renewcommand{\thepage}{S\arabic{page}}
\renewcommand{\theequation}{S\arabic{equation}}
\renewcommand{\thefigure}{S\arabic{figure}}
\renewcommand{\thetable}{S\arabic{table}}
\renewcommand{\thesection}{S\arabic{section}}
\renewcommand{\thesubsection}{S\arabic{section}.\arabic{subsection}}

\newpage
\section{Convergence of the IBI algorithm}

\begin{figure*}[h]
    \centering
    \includegraphics[width=.85\textwidth]{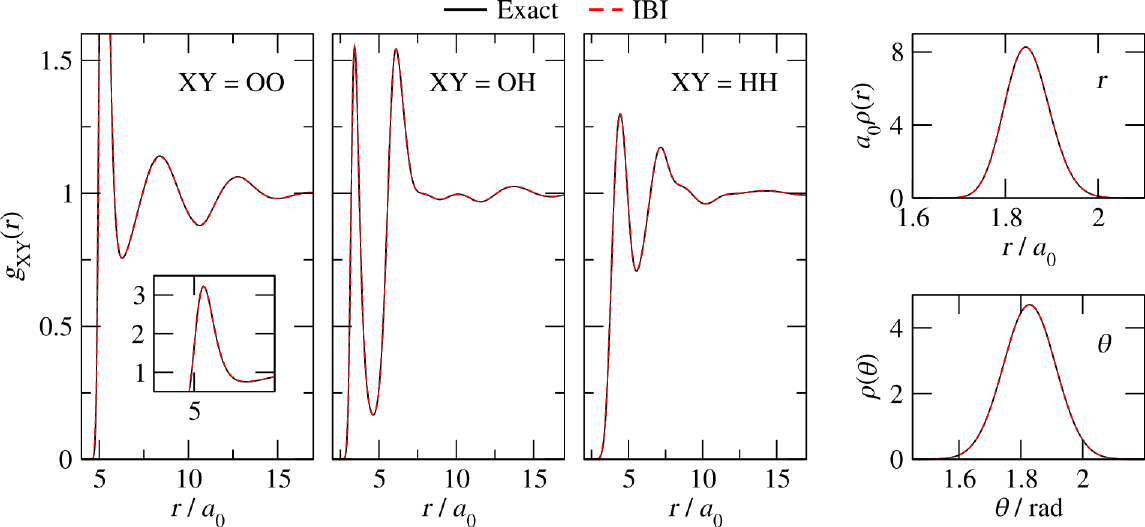}
    \caption{Comparison of distribution functions in the final IBI iteration to the exact distribution functions from a PIMD calculation for liquid water at 300\,K. (This is the same as Fig.~2 in the paper but we have included it here to facilitate the comparison with the ice results in Fig.~\ref{sifig:ibiconvergence-150K})}
    \label{sifig:ibiconvergence-300K}
\end{figure*}

\begin{figure*}[h]
    \centering
    \includegraphics[width=.85\textwidth]{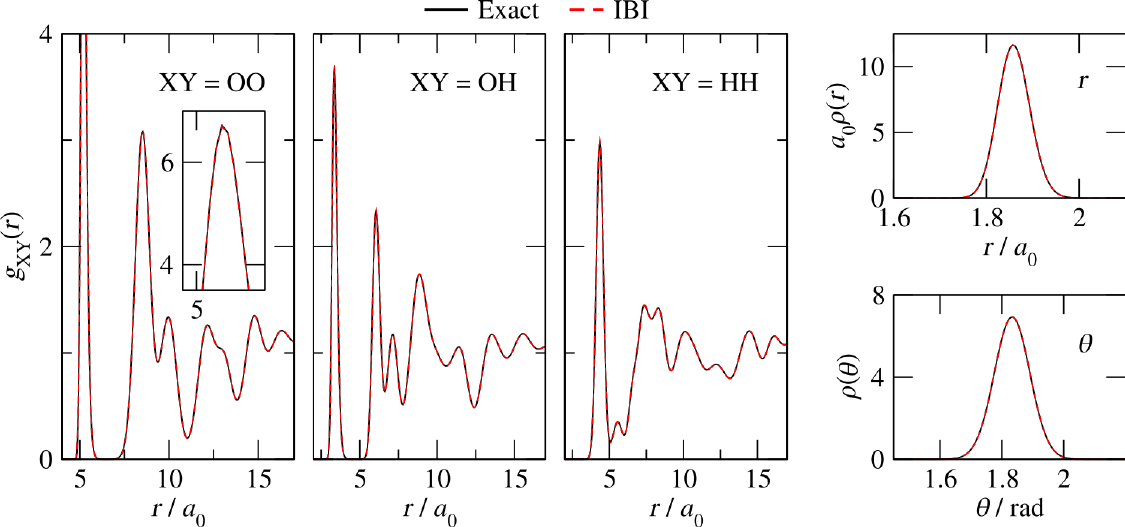}
    \caption{As figure \ref{sifig:ibiconvergence-300K}, but for hexagonal ice at 150\,K. }
    \label{sifig:ibiconvergence-150K}
\end{figure*}
\clearpage

\section{Quasi-centroid distribution functions}

\begin{figure*}[hb]
    \centering
    \includegraphics[width=.9\textwidth]{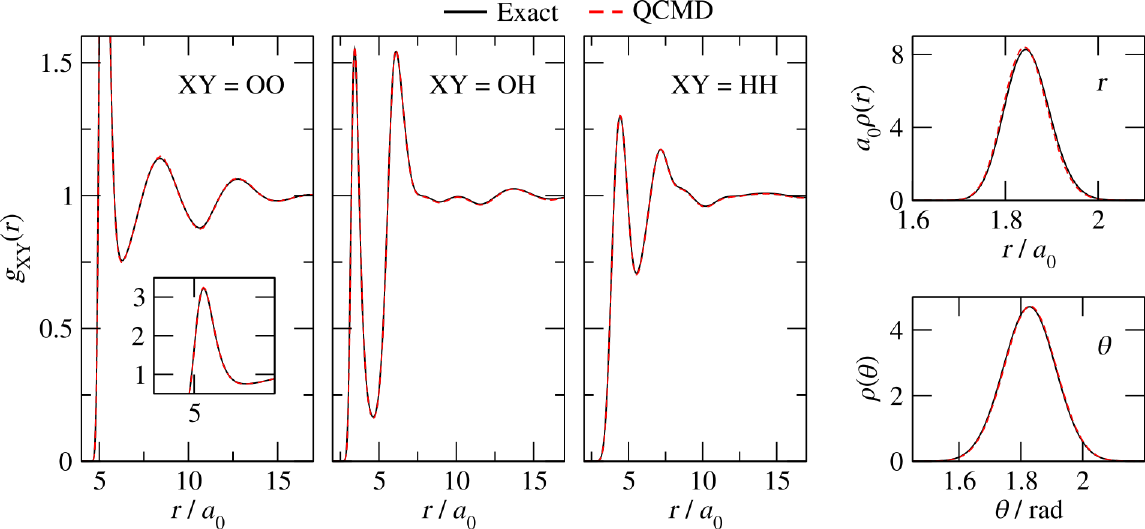}
    \caption{Quasi-centroid distribution functions averaged over a PIMD trajectory (black) and an adiabatic QCMD trajectory (red, dashed) for liquid water at 300\,K.}
    \label{sifig:quasicentroid-300K}
\end{figure*}

\begin{figure*}[ht]
    \centering
    \includegraphics[width=.9\textwidth]{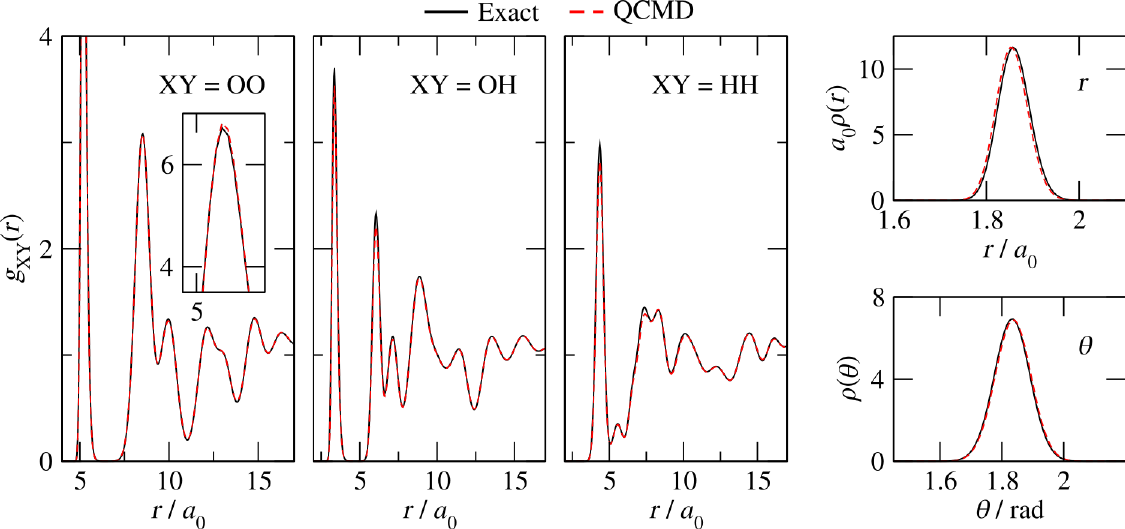}
    \caption{As Fig. \ref{sifig:quasicentroid-300K}, but for hexagonal ice at 150\,K.}
    \label{sifig:quasicentroid-150K}
\end{figure*}



